\documentclass[a4paper,11pt]{article}
\pdfoutput=1 % if your are submitting a pdflatex (i.e. if you have images in pdf, png or jpg format)
\usepackage{jheppub} % for details on the use of the package, please see the JHEP-author-manual
\usepackage[T1]{fontenc} % if needed
\usepackage[utf8]{inputenc}
\usepackage{braket}
\usepackage{cancel}
\usepackage[compat=1.1.0]{tikz-feynman} % for feynmann diagram
\usepackage{mathtools} % for dcase

\usepackage{subcaption} % for subfigures
\usepackage{cleveref} % for easier referincing
\usepackage{slashed} % Feynman slash
\usepackage{physics}
\usepackage{placeins} %such that figures are in the section they belong to
\usepackage[normalem]{ulem}

\definecolor{Orange}{cmyk}{0,0.61,0.87,0}
\definecolor{JungleGreen}{cmyk}{0.99,0,0.52,0}
\definecolor{OliveGreen}{cmyk}{0.64,0,0.95,0.40}
\definecolor{Brown}{cmyk}{0,0.81,1,0.60}
\definecolor{RoyalBlue}{cmyk}{0.71,0.53,0,0.12}
\definecolor{Gray}{cmyk}{0,0,0,0.40}
\definecolor{LightPink}{cmyk}{0.0,0.25,0,0}
\definecolor{LLightPink}{cmyk}{0.0,0.10,0,0}
\definecolor{LightBlue}{cmyk}{0.25,0,0,0}
\definecolor{LightGray}{cmyk}{0,0,0,0.2}
\definecolor{LightGreen}{cmyk}{0.3,0,0.3,0}
\def\l{\left}
\def\r{\right}

\def\be{\begin{equation}}
\def\ee{\end{equation}}
\def\bea{\begin{eqnarray}}
\def\eea{\end{eqnarray}}

\newcommand{\mDm}{m_\chi}
\newcommand{\g}[1]{\gamma_{#1}} % gamma matrix
\newcommand{\Lag}{\mathcal{L}} % Lagrangian
\newcommand{\hc}{\text{h.c.}} % hermitian conjugate

\newcommand{\fm}{\,\text{fm}}
\newcommand{\gev}{\,\text{GeV}} % GeV
\graphicspath{{./figures/}}
\title{\boldmath Anapole Moment of Majorana Fermions and Implications for Direct Detection of Neutralino Dark Matter}
\author[a]{Alejandro Ibarra}
\author[a]{Merlin Reichard}
\author[b]{Ryo Nagai}
\affiliation[a]{Physik-Department, Technische Universit\"at M\"unchen,\\
James-Franck-Stra\ss{}e, 85748 Garching, Germany}
\affiliation[b]{Department of Physics, Osaka University, Toyonaka, Osaka 560-0043, Japan}
\emailAdd{
ibarra@tum.de,
m.reichard@tum.de,
nagai@het.phys.sci.osaka-u.ac.jp}
\abstract{For Majorana fermions the anapole moment is the only allowed electromagnetic multipole moment. In this work we calculate the anapole moment induced at one-loop by the Yukawa and gauge interactions of a Majorana fermion, using the pinch technique to ensure the finiteness and gauge-invariance of the result. As archetypical example of a Majorana fermion, we calculate the anapole moment for the lightest neutralino in the Minimal Supersymmetric Standard Model, and specifically in the bino, wino and higgsino limits. Finally, we briefly discuss the implications of the anapole moment for the direct detection of dark matter in the form of Majorana fermions. }
\begin{document} 
\maketitle
\flushbottom
%%%%%%%%%%%%%%%%%%%%

\section{Introduction} 

There is mounting evidence for the existence of dark matter in galaxies, clusters of galaxies and the Universe at large scale, possibly in the form of a population of new elementary particles, not contained in the Standard Model of Particle Physics  (for reviews, see {\it e.g.} \cite{Jungman:1995df,Bertone:2004pz,Bergstrom:2000pn,Feng:2010gw}). Astronomical and cosmological observations demonstrate that dark matter particles interact with the electromagnetic radiation much more weakly than the hadrons or the charged leptons, however observations do not require the dark matter particle to be completely decoupled from the photon. 

As is well known, for an electrically neutral fermion, the Lorentz- and gauge symmetries allow a magnetic- and electric dipole moment, and an anapole moment~\cite{Fujikawa:1980yx,Petcov:1976ff,Pal:1981rm,Shrock:1982sc,Giunti:2008ve, Nieves:1981zt,Kayser:1982br}.  For an electrically neutral complex vector, also electric and magnetic quadrupole moments exist in general ~\cite{Aronson:1969ltq,Gaemers:1978hg,Hagiwara:1986vm}. Therefore, even if the dark matter particle is electrically neutral, it may couple to the photon via the different electromagnetic multipoles. In fact, if the dark matter particle has interactions with the Standard Model particles, as many models suggest, such electromagnetic multipoles will be generically generated at the quantum level. This interaction could play an important role in the direct detection of dark matter particles through the scattering with nuclei, as discussed in several works, {\it e.g.} \cite{Pospelov:2000bq,Sigurdson:2004zp,Masso:2009mu,Chang:2010en,Barger:2010gv,Banks:2010eh,Ho:2012bg,DelNobile:2012tx,Kopp:2014tsa,Ibarra:2015fqa,Hambye:2021xvd,Hisano:2020qkq,DelNobile:2014eta,Garny:2015wea,Cabral-Rosetti:2015cxa,Abe:2018emu,Sandick:2016zut,Kang:2018oej,Baker:2018uox,Arina:2020mxo,Kuo:2021mtp}. 

In this paper we will concentrate on Majorana fermions as dark matter candidates. In this case, the invariance of the Majorana field under the charge conjugation operation only allows the anapole moment. Concretely, we will consider the lightest neutralino in the Minimal Supersymmetric Standard Model (MSSM) as an archetype of Majorana dark matter. The gauge and Yukawa interactions of the lightest neutralino with the charged particles of the Standard Model will generate an anapole moment at the one loop level. In the pure bino limit, the dark matter particle is a singlet Majorana fermion that couples to the Standard Model fermions via a t-channel mediator (the sfermions). The anapole moment of the singlet Majorana fermion has been calculated in \cite{Kopp:2014tsa}. In the pure wino and higgsino limits (as well as in the mixed cases), the dark matter particle has $SU(2)_L$ charge and also interacts with the $W$-boson, and special care has to be taken in order to ensure the gauge invariance of the result. 

Similar challenges have been found in the past in the calculation of the neutrino charge radius \cite{Bardeen:1972vi,Abers:1973qs,Fujikawa:1972fe,Lee:1973fw,PhysRevD.43.2956}, or in the calculation of the off-shell magnetic form factors of quarks and leptons in non-Abelian gauge theories  \cite{Papavassiliou:1993qe,CabralRosetti:2002tf,Fujikawa:1972fe,Bernabeu:2007rr}, which naively yield gauge-dependent results. The gauge dependence, clearly unphysical, arises due to redundancies in the individual Green functions which  are introduced by the gauge-fixing procedure. This problem was solved by introducing the so-called  pinch technique \cite{Cornwall:1989gv,Papavassiliou:1989zd,Bernabeu:2000hf,PhysRevD.61.013001,Bernabeu:2002pd}, which consists in an algorithmic diagrammatic construction of physical subamplitudes by resumming topologically similar terms within an amplitude and ultimately defining a proper and physical vertex by including only vertex-like contributions (see \cite{Binosi:2009qm} for a review). Furthermore, the resulting effective Green functions coincides with the one calculated using the background field method in the quantum Feynman gauge \cite{Denner:1994nn,Hashimoto:1994ct,Papavassiliou:1994yi}.

In this paper we will apply the background field method to calculate the anapole moment of a spin 1/2 Majorana fermion that interacts both via Yukawa couplings and via gauge couplings, showing explicitly that the result is finite and gauge invariant. The general result is presented in \cref{sec:1-loop_calculation}. In  \cref{sec:AnapoleMoment_MSSM}  we particularize our results to the lightest neutralino in the MSSM, and in \cref{sec:SimplifiedMSSM} we study some well motivated MSSM scenarios. Then, in \cref{sec:DDlimits_pMSSM} we briefly discuss the prospects of detection of a Majorana dark matter candidate via its anapole moment in direct search experiments, and finally in \cref{sec:conclusions} we present our conclusions. We also include Appendix \ref{sec:ParticleSpectrumMSSM} summarizing the calculation of the particle mass eigenstates in terms of the interaction eigenstates in the MSSM.

%%%%%%%%%%%%%%%%%%%%%%%%%%%%
\section{One-Loop Calculation of the Anapole Moment of a Majorana Fermion}\label{sec:1-loop_calculation}

We consider a Majorana fermion, that we denote by $\chi$, with mass $m_\chi$. The interaction vertex of a Majorana fermion with the photon is restricted by the Lorentz- and gauge symmetries to be of the form \cite{Nieves:1981zt,Kayser:1982br,Giunti:2008ve}
\begin{equation}
M_\mu(q)=f_A(q^2) (q^2\g\mu -q_\mu \slashed{q})\g5,
\end{equation}
where $q_\mu$ denotes the photon outgoing momentum and $f_A(q^2)$ is the anapole form factor. This interaction vertex generates at low momentum the C- and P-violating effective Lagrangian
\begin{equation}\label{eq:EffectiveLagrangianAnapole}
\Lag_\text{eff} = \frac{\mathcal{A}}{2}\, \bar{\chi} \g\mu\g5\chi\partial_\nu F^{\mu\nu}\;,
\end{equation}
where $\mathcal{A}$ is the anapole moment, defined as the zero momentum limit of the anapole form factor:
\begin{equation}\label{eq:definitionAnapoleMoment}
\mathcal{A}=\lim\limits_{q^2\rightarrow 0} f_A(q^2).
\end{equation}

Being $\chi$ electrically neutral, the interaction with the photon can only arise at the quantum level through a coupling with charged particles. In this work, we will consider the cases where $\chi$ interacts with a charged gauge boson and/or with a charged scalar. 

We first consider a scenario where $\chi$ couples to a charged Dirac fermion,  $\chi^\mp$ with mass $m_{\chi^-}$, and an electrically charged gauge boson, $V^\pm$ with mass $m_V$ acquired through the spontaneous breaking of a gauge symmetry. The interaction Lagrangian reads:
\begin{align}
\label{eq:InteractionsLagrangian:Vector}
	\Lag_\text{FFV}	&= \bar\chi\gamma^\mu \left[v_L P_L + v_R P_R\right] \chi^- V_\mu^+ %MR 16.02.2022 changed convention to Ryo's,
	+\bar\chi\left[c^G_L P_L + c^G_R P_R\right]  \chi^- G^+ +\hc,
\end{align}
where $G^\pm$ are the Goldstone bosons arising from the spontaneous symmetry breaking. 

To calculate the anapole moment we employ the background field method (BFM) \cite{Denner:1994xt,Denner:1994nn}. Namely, we replace the photon with the background photon $\gamma\rightarrow \hat\gamma$ and the $\gamma VV$-vertex with its BFM version in the quantum Feynman gauge. Explicitly, the triple gauge vertex reads \cite{Denner:1994xt}
\begin{align}\label{eq:BFMFeynmanRulesGaugeVerticesAWW}
	 i \hat\Gamma_{\gamma VV}^{\mu \nu\rho}(k_1,k_2,k_3) &= -i e \left[g_{\nu\rho}(k_3-k_2)_\mu + g_{\mu\nu}(k_2-k_1+k_3)_\rho +g_{\rho\mu} (k_1-k_3-k_2)_\nu \right],
\end{align}
while the gauge-gauge-Goldstone vertex reads:
\begin{align}\label{eq:BFMFeynmanRulesGaugeVerticesAWG}
	 i\hat\Gamma_{\gamma VG}^{\mu \nu}(k_1,k_2,k_3) &= 0.
\end{align}

\begin{figure}
	\centering
	\begin{subfigure}[c]{1\linewidth}
%		\begin{equation*}	
%		\feynmandiagram[small,vertical=i3 to e2] {
%			i1 --[ boson, edge label' = \( V \)] i2 --[fermion, edge label' = \( \chi^- \)] i3 --[fermion, edge label' = \( \chi^- \)] i1 -- a[particle=\(\chi\)],
%			e1 [particle=\(\chi\)] -- i2,
%			e2 [particle=\({\hat\gamma}\)] --[photon] i3,
%		};
%		\feynmandiagram[small, vertical= i3 to e2] {
%			i1 --[anti fermion, edge label' = \( \chi^- \)] i2 --[ boson, edge label' = \( V \)] i3 --[boson, edge label' = \( V \)] i1 -- a[particle=\(\chi\)],
%			e1 [particle=\(\chi\)] -- i2,
%			e2 [particle=\({\hat\gamma}\)] --[photon] i3,
%		};
%		\feynmandiagram[small, vertical= i3 to e2] {
%		i1 --[charged scalar, edge label' = \( G^-\)] i2 --[fermion, edge label' = \( \chi^- \)] i3 --[fermion, edge label' = \( \chi^- \)] i1 -- a[particle=\(\chi\)],
%		e1 [particle=\(\chi\)] -- i2,
%		e2 [particle=\({\hat\gamma}\)] --[photon] i3,
%	};	
%		\feynmandiagram[small, vertical= i3 to e2] {
%		i1 --[anti fermion , edge label' = \( \chi^- \)] i2 --[anti charged scalar, edge label' = \( G^- \)] i3 --[anti charged scalar, edge label' = \( G^-  \)] i1 -- a[particle=\(\chi\)],
%		e1 [particle=\(\chi\)] -- i2,
%		e2 [particle=\({\hat\gamma}\)] --[photon] i3,
%	};	
%		\end{equation*}
\includegraphics{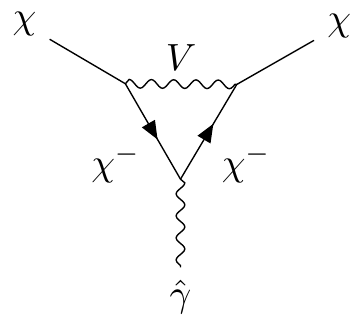}
\includegraphics{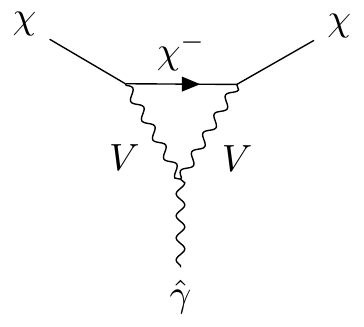}
\includegraphics{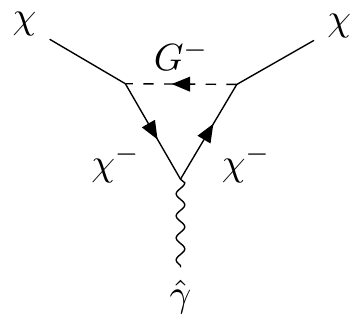}
\includegraphics{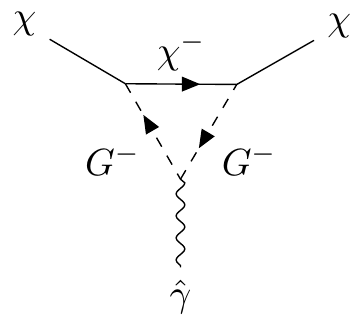}
	\end{subfigure}
\centering
	\begin{subfigure}[c]{1\linewidth}
        \centering
%		\begin{equation*}
%		\feynmandiagram[small, vertical= i3 to e2] {
%			i1 --[charged scalar, edge label' = \( S\)] i2 --[fermion, edge label' = \( f \)] i3 --[fermion, edge label' = \( f \)] i1 -- a[particle=\(\chi\)],
%			e1 [particle=\(\chi\)] -- i2,
%			e2 [particle=\({\hat\gamma}\)] --[photon] i3,
%		};		
%		\feynmandiagram[small, vertical= i3 to e2] {
%			i1 --[anti fermion , edge label' = \( f \)] i2 --[anti charged scalar, edge label' = \( S \)] i3 --[anti charged scalar, edge label' = \( S  \)] i1 -- a[particle=\(\chi\)],
%			e1 [particle=\(\chi\)] -- i2,
%			e2 [particle=\({\hat\gamma}\)] --[photon] i3,
%		};		
%		\end{equation*}
\includegraphics{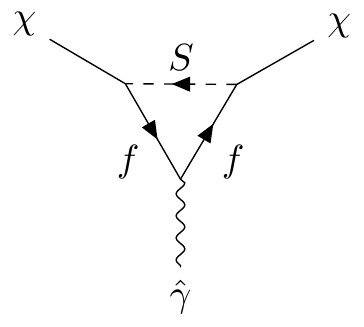}
\includegraphics{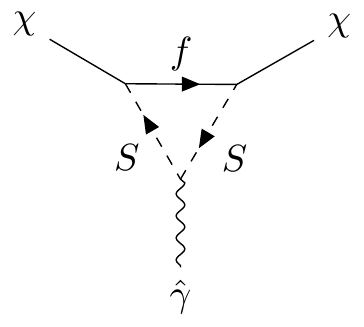}
	\end{subfigure}
	\caption{Feynman diagrams generating an anapole moment for a Majorana fermion at the one-loop level in the background field formulation, via the mediation of an electrically charged vector boson (top) or the mediation of a charged scalar (bottom). 
	}
	\label{fig:TriangleDiagramVectorandScalar}
\end{figure}

\begin{figure}[t!]
	\centering
		\includegraphics[width=.49\textwidth, clip = true]{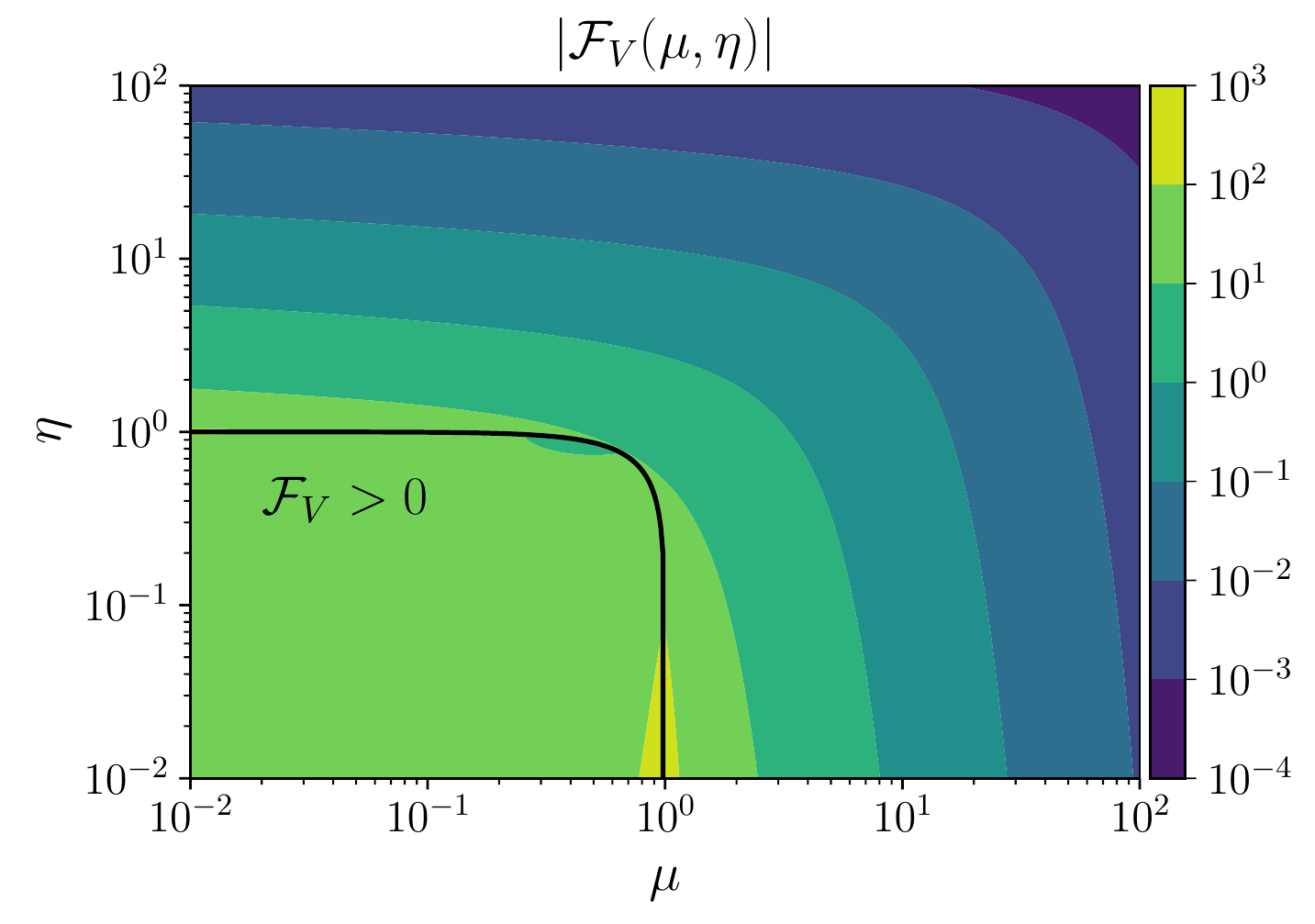}
		\includegraphics[width=.49\textwidth,clip = true]{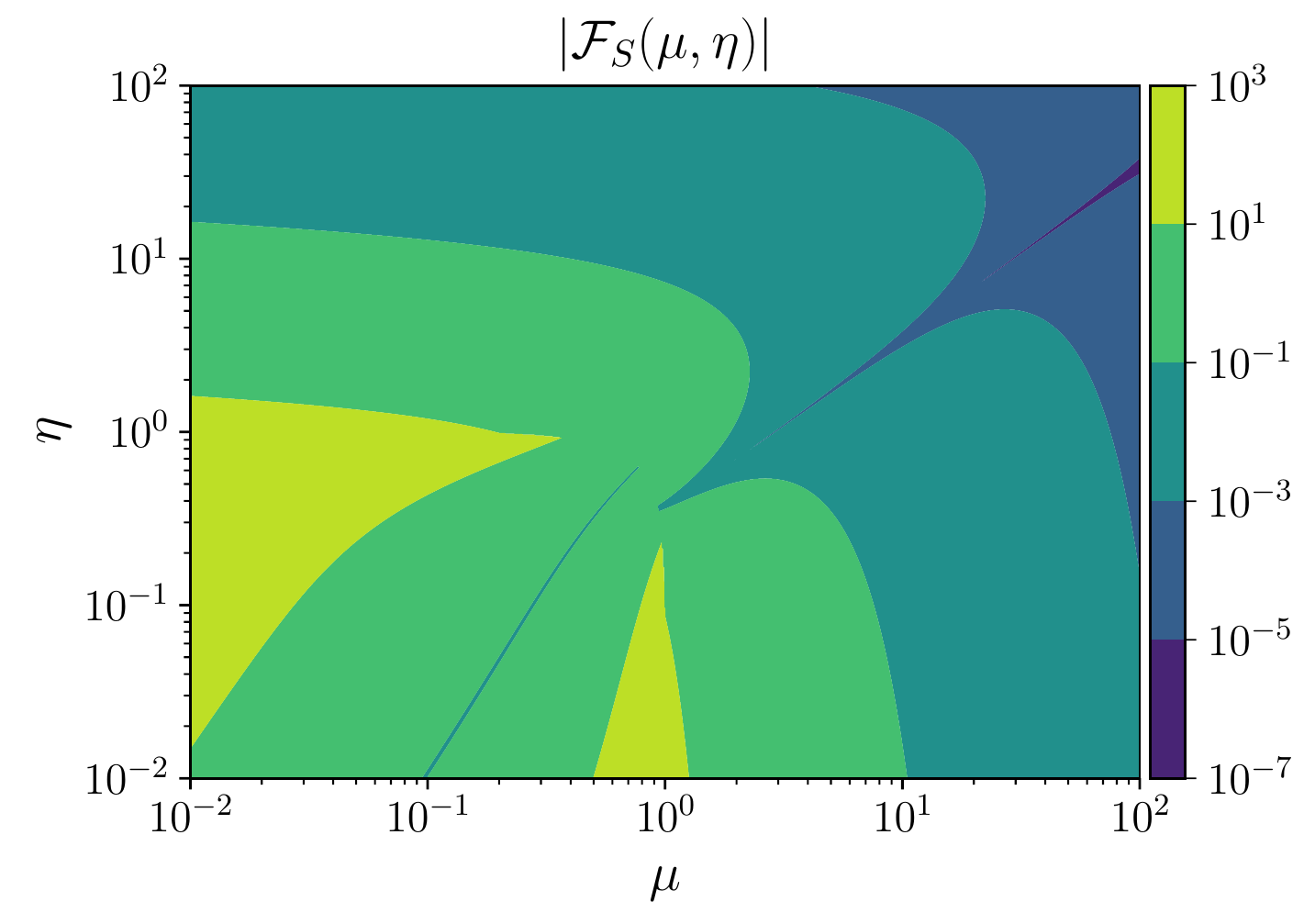}
	\caption{Absolute values of the functions $\mathcal{F}_V(\mu,\eta)$ (left) and $\mathcal{F}_{S}(\mu,\eta)$ (right).}
	\label{fig:Fscalar_Fvector_Scan}
\end{figure}

The one-loop diagrams relevant for the calculation of the anapole moment in the BFM are shown in  \cref{fig:TriangleDiagramVectorandScalar}. We obtain\footnote{
We have used the Feynman rules for Majorana fermions derived in \cite{Denner:1992vzaFeynmanRulesMajorana1,Denner:1992meFeynmanRulesMajorana2}; the calculation was performed with the help of \texttt{FeynCalc} \cite{MERTIG1991345,Shtabovenko:2016sxi,Shtabovenko:2020gxv}, \texttt{FeynHelpers} \cite{Shtabovenko:2016whf} and \texttt{Package-X} \cite{Patel:2015tea}.}:
\begin{align}\label{Anapole:eq:Fvector}
\mathcal{A}_V
&=\frac{e}{96\pi^2\mDm^2}\Big\{2\left[\abs{v_L}^2-\abs{v_R}^2\right]\mathcal{F}_V\Big(\frac{m_{\chi^-}}{m_\chi},\frac{m_V}{m_\chi}\Big)+  \left[\abs{c^G_L}^2-\abs{c^G_R}^2\right]\mathcal{F}_S\Big(\frac{m_{\chi^-}}{m_\chi},\frac{m_V}{m_\chi}\Big)\Big\},
\end{align}
where 
\begin{align}
\mathcal{F}_X(\mu,\eta)&=\frac{3}{2}\log(\frac{\mu^2}{\eta^2})+(3\eta^2-3\mu^2+n_X) f(\mu,\eta),
\end{align}
for $X=V,S$, with $n_V=-7$, $n_S=1$, 
and
\begin{align}\label{eq:loopFunctionf}
f(\mu,\eta)&=\begin{cases} 
\frac{1}{2\sqrt\Delta} \log\frac{\mu^2+\eta^2-1+\sqrt{\Delta}}{\mu^2+\eta^2-1-\sqrt{\Delta}} & \Delta\neq 0 \\
\frac{2}{(\mu^2-\eta^2)^2-1} & \Delta=0
\end{cases}\;,
\end{align}
with $\Delta\equiv\Delta(\mu,\eta)=(\mu^2+\eta^2-1)^2-4\mu^2\eta^2$.  Contour plots of $\mathcal{F}_{V}(\mu,\eta)$ and $\mathcal{F}_{S}(\mu,\eta)$  are shown in  \cref{fig:Fscalar_Fvector_Scan}, and present a  discontinuity at $\mu^2+\eta^2=1$. The anapole interaction, being P-violating, must vanish if the underlying model preserves parity, namely when $v_L=v_R$ and $c^G_L=c^G_R$, as apparent from Eq.~(\ref{Anapole:eq:Fvector}).

We also consider the scenario where  $\chi$ couples to a charged Dirac fermion $f$ and a charged complex scalar $S$, with masses $m_f$ and $m_S$, and with charges $\pm e Q_f$ respectively (see also  \cite{Kopp:2014tsa,Garny:2015wea,Sandick:2016zut,Baker:2018uox}).
The interaction Lagrangian in this case can be written as
\begin{align}
\label{eq:InteractionsLagrangian:Scalar}
	\Lag_\text{FFS} &= \bar\chi\left[c_L P_L + c_R P_R\right] S^* f +\hc,
\end{align}
allowing $\chi$ to interact with the (background) photon field via the diagrams shown in  \cref{fig:TriangleDiagramVectorandScalar}.
The induced scalar contribution to the anapole moment reads
\begin{equation}\label{Anapole:eq:Fscalar}
	\mathcal{A}_S =- \frac{e}{96\pi^2\mDm^2}Q_f\left[\abs{c_L}^2-\abs{c_R}^2\right]\mathcal{F}_S\Big(\frac{m_f}{m_\chi},\frac{m_S}{m_\chi}\Big)\;.
\end{equation}

\begin{figure}[!hbt]
	\centering
	\includegraphics[width=1\linewidth]{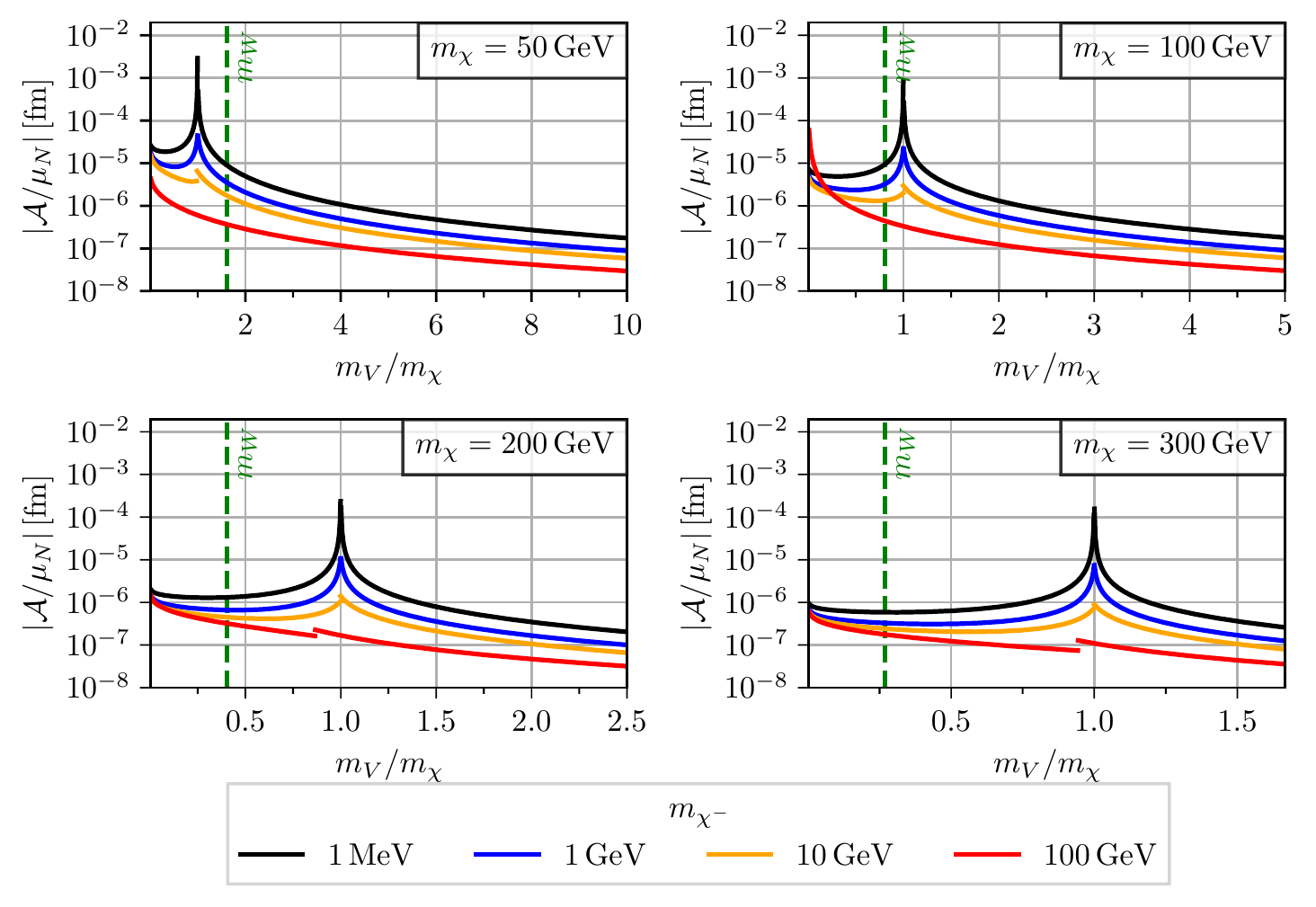}
	\caption{Anapole moment (normalized to the nuclear magneton $\mu_N$)  induced at the one loop level via the interaction of the Majorana fermion $\chi$ with a charged gauge boson $V$ and a charged fermion $\chi^-$, as a function of $m_V/m_\chi$, for different values of the Majorana mass and different values of the charged fermion mass. For the plot, we assumed for concreteness $v_L=1$ and $v_R, c_L^G, c_R^G=0$.}
	\label{fig:VectorAnapolePlot}
\end{figure}

\begin{figure}[!hbt]
	\centering
	\includegraphics[width=1\linewidth]{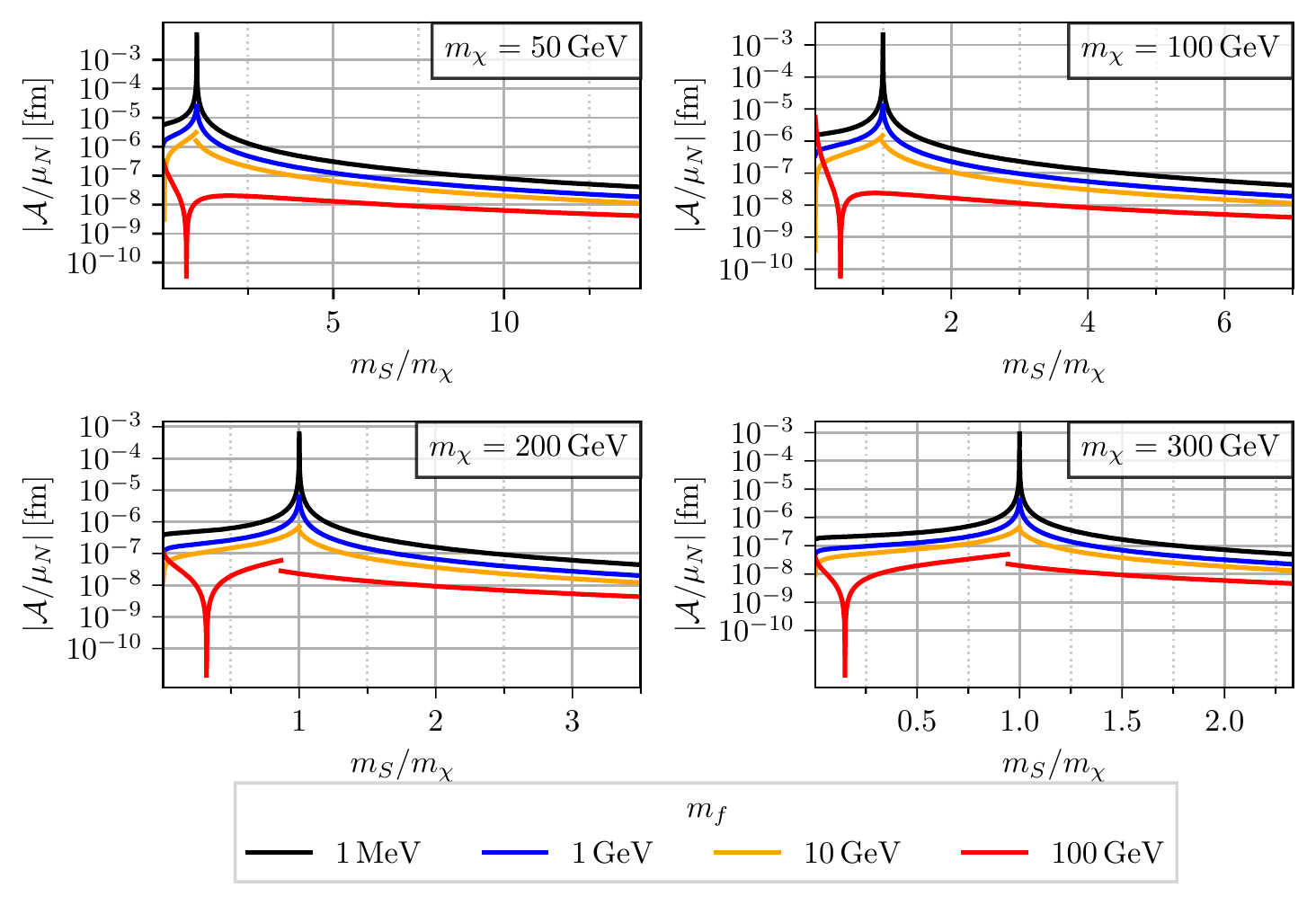}
	\caption{Same as Fig. \ref{fig:VectorAnapolePlot} but for a Majorana fermion that interacts with a charged scalar $S$ and a charged fermion $f$.  For the plot, we assumed for concreteness $c_L=1$, $c_R=0$ and $Q_f=-1$.}
	\label{fig:ScalarAnapolePlot}
\end{figure}

The resulting vector- and scalar contributions to the anapole moment of $\chi$ are shown in \cref{fig:VectorAnapolePlot,fig:ScalarAnapolePlot} as a function of $\eta=m_{S,V}/\mDm$, for different values of the fermion mass $m_{f, \chi^-}$. In both cases the anapole moment is enhanced for $\eta\approx 1$ and $m_{f,\chi^-} \ll \mDm$.
The dependence of the anapole moment with $\mu=m_{f,\chi^-}/\mDm$ is similar, see also \cref{fig:Fscalar_Fvector_Scan}.  \footnote{The ``compressed'' spectrum  requires a certain adjustment of the fundamental parameters of the model. On the other hand, it is a viable possibility from the phenomenological point of view, and has attracted some attention in the literature in the context of dark matter production via coannihilations~\cite{Griest:1990kh,Baker:2015qna}, indirect detection~\cite{Flores:1989ru,Garny:2011ii}, or collider searches~\cite{Martin:2007gf,Dreiner:2012gx}. For an extensive review, see~\cite{Garny:2015wea}. }

The general formulas \cref{Anapole:eq:Fvector} and \cref{Anapole:eq:Fscalar} simplify when the charged fermions are much heavier than the vector/scalar and the Majorana fermion, {\it i.e.} when $m_{f,\chi^-}\gg m_{S,V}, m_\chi$. In this case,   we find
\begin{align}\label{eq:AnapoleFunctionsHeavyeInternalFermionLimit}
    \mathcal{A}_V & \simeq  \frac{e}{96\pi^2m_{\chi^-}^2}\left\{2\left[\abs{v_L}^2-\abs{v_R}^2\right] \left(3+5\log{\frac{m_V^2}{m_{\chi^-}^2}}\right)+\left[\abs{c_L^G}^2-\abs{c_R^G}^2\right] \left(3+\log{\frac{m_V^2}{m_{\chi^-}^2}}\right)\right\}, \nonumber \\
    \mathcal{A}_S &\simeq  -\frac{e}{96\pi^2m_f^2}Q_f\left[\abs{c_L}^2-\abs{c_R}^2\right] \left(3+\log{\frac{m_S^2}{m_f^2}}\right),
\end{align}
which are independent of $m_\chi$. Analogously, in the heavy scalar/vector limit, {\it i.e} when  $m_{V,S}\gg m_\chi, m_f$ 
\begin{align}
    \mathcal{A}_V &\simeq  \frac{e}{96\pi^2m_V^2}\left\{2\left[\abs{v_L}^2-\abs{v_R}^2\right] \left(-3+2\log{\frac{m_{\chi^-}^2}{m_V^2}}\right) - \left[\abs{c_L^G}^2-\abs{c_R^G}^2\right] \left(3+2\log{\frac{m_{\chi^-}^2}{m_V^2}}\right) \right\},\nonumber \\ \label{eq:AnapoleFunctionsHeavyeInternalBosonLimit}
    \mathcal{A}_S &\simeq  \frac{e}{96\pi^2m_S^2}Q_f\left[\abs{c_L}^2-\abs{c_R}^2\right] \left(3+2\log{\frac{m_f^2}{m_S^2}}\right).
\end{align}

This general formalism can be applied in particular to calculate the anapole moment of the Standard Model neutrinos, through its interactions with the $W$ boson and the charged leptons. In this case, $v_L=g/\sqrt{2}$, $v_R=c_L^G=c_R^G=0$, resulting in~ \cite{Bernabeu:2000hf,Bernabeu:2002pd}
\begin{align}
{\cal A}&
\simeq \frac{e G_F}{12 \sqrt{2} \pi^2} (-3+2\log\frac{m_\ell^2}{m_W^2}),
\end{align}
where $G_F=\sqrt{2} g^2/8 m_W^2$ is the Fermi constant, with $m_W$ the W-boson mass, and $m_\ell$ is the charged lepton mass.

For Dirac fermions, the anapole moment is half as large as for Majorana fermions, due to the halving of the number of diagrams.

\section{Anapole Moment of the Lightest Neutralino in the MSSM}\label{sec:AnapoleMoment_MSSM}

An archetype of Majorana fermion interacting with charged particles both via a charged vector mediator and a charged scalar mediator is the lightest neutralino in the MSSM. The MSSM Lagrangian contains an interaction term between the lightest neutralino, the charginos $\chi_j$, $j=1,2$, and the W-boson (and its Goldstone boson). This term has the form
\begin{align}
\label{eq:InteractionsLagrangian:VectorMSSM}
	\Lag	&\supset \bar\chi\gamma^\mu \left[v^j_L P_L + v^j_R P_R\right] \chi_j^- W_\mu^+
	+\bar\chi\left[c^{G,j}_L P_L + c^{G,j}_R P_R\right]\chi_j^- G^+ +\hc,
\end{align}
with
\begin{align}\label{eq:MSSMVectorCouplings}
v^j_L &= -gN_{12} U_{j1}^* - g\frac{1}{\sqrt2} N_{13} U_{j2}^*,\nonumber \\
v^j_R &= -gN_{12}^* V_{j1}   +g\frac{1}{\sqrt2} N_{14}^* V_{j2}, \nonumber\\
c_L^{G,j} &=g\cos\beta\left[N_{13}^* U_{j1}^*-\frac{1}{\sqrt 2}U_{j2}^*(N_{12}^*+\tan\theta_W N_{11}^*)\right], \nonumber \\
c_R^{G,j} &=-g\sin\beta\left[N_{14}V_{j1}+\frac{1}{\sqrt 2} V_{j2}(N_{12}+\tan\theta_W N_{11})\right],
\end{align}
where $\tan\beta=\langle H_2^0\rangle/\langle H_1^0\rangle$ denotes the ratio between the expectation values of the neutral components of the up-type Higgs and the down-type Higgs doublet, $\theta_W$ is the Weinberg's angle, $N_{ij}$, $i,j=1,\ldots,4$ are the elements of the neutralino mixing matrix, and $U_{ij}$ ($V_{ij}$), $i,j=1,2$ are the elements of the mixing matrix of the negatively (positively) charged chargino (see \cref{sec:ParticleSpectrumMSSM} for a brief summary of the construction of the mass eigenstates in the MSSM from the interaction eigenstates).

Further, the Lagrangian contains an interaction term with the chargino and the charged Higgs, of the form
\begin{align}
\label{eq:InteractionsLagrangian:Higgs}
	\Lag_\text{FFS} &\supset \bar\chi\left[c^{H,j}_L P_L + c^{H,j}_R P_R\right] H^+ \chi_j^- +\hc,
\end{align}
with
\begin{align}
		c_L^{H,j} &= -g\sin\beta\left[N_{13}^* U_{j1}^*-\frac{1}{\sqrt 2}U_{j2}^*(N_{12}^*+\tan\theta_W N_{11}^*)\right], \nonumber \\
		c_R^{H,j} &=-g\cos\beta\left[N_{14}V_{j1}+\frac{1}{\sqrt 2} V_{j2}(N_{12}+\tan\theta_W N_{11})\right],
		\label{eq:couplings_chargino_chargedH_R_L}
\end{align}
as well as an interaction term with the SM fermions and sfermions of the form
\begin{align}
\label{eq:InteractionsLagrangian:Sfermions}
	\Lag_\text{FFS} &\supset  \bar\chi\left[c^{i,a}_L P_L + c^{i,a}_R P_R\right] {\widetilde f}_a f_i +\hc,
\end{align}
with
\begin{align}
c^{i,1}_L&=G^{{f}_{iL}}\cos\theta_{\widetilde{f}_a} + H^{{f}_{iR}} \sin\theta_{\widetilde{f}_a}, \nonumber \\
c^{i,1}_R&=G^{{f}_{iR}}\sin\theta_{\widetilde{f}_a} + H^{{f}_{iL}} \cos\theta_{\widetilde{f}_a}, \nonumber \\
c^{i,2}_L&=-G^{{f}_{iL}}\sin\theta_{\widetilde{f}_a} + H^{{f}_{iR}} \cos\theta_{\widetilde{f}_a}, \nonumber \\
c^{i,2}_R&=G^{{f}_{iR}}\cos\theta_{\widetilde{f}_a} - H^{{f}_{iL}} \sin\theta_{\widetilde{f}_a},
\end{align}
and 
\begin{align}
G^{{f}_{iL}}&=-\sqrt{2}g\left[T^{f_i}_{3L}N^*_{12}+\tan\theta_W(Q_{f_i}-T^{f_i}_{3L})N^*_{11}\right], \nonumber \\
G^{{f}_{iR}}&=\sqrt{2}g\tan\theta_WQ_{f_i} N_{11}, \nonumber\\
H^{{f}_{iL}} &=- \frac{g}{\sqrt{2}m_W}m_{f_i} \times \begin{cases}
	N_{14}/\sin\beta, & f_i =u\text{-type} \nonumber\\
	N_{13}/\cos\beta, & f_i = d\text{-type}, \ell\\
\end{cases}\\
H^{{f}_{iR}} &= H^{{f}_{iL}*}.
\end{align}

These interactions induce at the one loop level an anapole moment for the lightest neutralino $\chi$:
\begin{align}
   \mathcal{A}=\mathcal{A}_W+\mathcal{A}_{\widetilde{f}}+\mathcal{A}_H.
 \end{align}
Using the general results of \cref{sec:1-loop_calculation} for the vector and scalar contributions to the anapole moment of a Majorana fermion, one finds
 \begin{align}
    \mathcal{A}_W &= \frac{e}{96\pi^2\mDm^2}
    \Big\{2\sum_{j}\left[|v^{j}_L|^2-|v^{j}_R|^2\right]\mathcal{F}_W\Big(\frac{m_{\chi^-_j}}{\mDm},\frac{m_{W^+}}{\mDm}\Big)\nonumber \\ \nonumber
    &~~~~~~~~~~~~~~~+\sum_{j}\left[|c^{G,j}_L|^2-|c^{G,j}_R|^2\right]\mathcal{F}_S\Big(\frac{m_{\chi^-_j}}{\mDm},\frac{m_{W^+}}{\mDm}\Big)\Big\}, \nonumber\\
    \mathcal{A}_{\widetilde{f}} &= - \frac{e}{96\pi^2\mDm^2}\sum_{i,a}N^i_cQ_i\left[|c^{i,a}_L|^2-|c^{i,a}_R|^2\right]\mathcal{F}_S\Big(\frac{m_{f_i}}{\mDm},\frac{m_{\widetilde{f}_a}}{\mDm}\Big), \nonumber \\
    \mathcal{A}_{H} &=- \frac{e}{96\pi^2\mDm^2}\sum_{j}Q_j\left[|c^{H,j}_L|^2-|c^{H,j}_R|^2\right]\mathcal{F}_S\Big(\frac{m_{\chi^-_j}}{\mDm},\frac{m_{H^+}}{\mDm}\Big).
\end{align}

In what follows, we will particularize these expressions to some well motivated MSSM scenarios. 

\section{MSSM Scenarios}\label{sec:SimplifiedMSSM}

In order to gain insight into the rich physics of supersymmetric models, we will study first in subsections  \ref{sec:pure_heavy_sfermion}, \ref{sec:Light_sfermions} and
\ref{sec:mixed_heavy_sfermion} some simplified scenarios where some SUSY particles are integrated out. Lastly, in subsection 
\ref{sec:FullMSSM}, we will consider a general MSSM scenario.

\subsection{Pure lightest neutralino \& heavy sfermions}
\label{sec:pure_heavy_sfermion}

Let us first consider a number of toy models where the lightest neutralino practically coincides with an interaction eigenstate, either the bino, the higgsino, or the wino. We also consider first that all sfermions are decoupled, so that ${\cal A}_{\widetilde f}\simeq 0$.

\subsubsection*{Bino limit}

In the limit $M_1\ll M_2, |\mu|, m_{\widetilde f}$, the lightest neutralino is practically inert and in particular does not couple to the $W$ boson, so that ${\cal A}_{W}\simeq 0$, nor to the charged Higgs, so that  ${\cal A}_{H}\simeq 0$. The anapole moment in this toy model is therefore expected to be very suppressed. 

\subsubsection*{Higgsino limit}

In the limit $|\mu|\ll M_1, M_2, m_{\widetilde f}$ the two lightest neutralinos are nearly degenerate in mass and form a pseudo-Dirac pair. Further, there is only one light chargino, which practically coincides with the charged Higgsino. The effective couplings of the lightest neutralino to the $W$-boson and the lightest chargino read:
\begin{align}
&v^1_L  \simeq -\frac{g}{2},~~~~~
v^1_R \simeq \frac{g}{2}, \nonumber \\
&c_L^{G,1} \simeq 0,~~~~~~
c_R^{G,1} \simeq 0.
\end{align}
and are manifestly parity conserving. The vector contribution to the anapole moment is therefore suppressed in this scenario, ${\cal A}_W\simeq 0$. The same result holds for a minimal dark matter scenario where the dark matter particle is a Majorana fermion, doublet under $SU(2)_L$ and with hypercharge $1/2$. 

In the MSSM, moreover, the lightest neutralino also couples to the chargino and to the charged Higgs, which may be light. On the other hand, it follows from \cref{eq:couplings_chargino_chargedH_R_L} that in the Higgsino limit the coupling strengths are 	$c_R^{H,1}, c_L^{H,1} \simeq 0$, and therefore ${\cal A}_{H}\simeq 0$ regardless of the mass of the charged Higgs.

\subsubsection*{Wino limit} 

In the limit $M_2\ll M_1, |\mu|, m_{\widetilde f}$, there is only one light neutralino and one light chargino, which are composed mainly by a neutral wino and a charged wino respectively. In this limit, the anapole moment only receives contributions from the chargino-$W$ loop. The relevant coupling constants read 
\begin{align}
&v^1_L = -g , ~~~~~
v^1_R = -g, \nonumber \\
&c_L^{G,1} =0,  ~~~~~~
c_R^{G,1} =0.
\end{align}
which are manifestly parity conserving and lead to a suppressed anapole moment.

It is apparent from these limiting cases that in order to enhance the anapole moment it is necessary to couple the lightest neutralino to new light particles with parity breaking interactions, and/or to introduce an admixture in the neutralino eigenstate of different interaction eigenstates. We discuss these two possibilities below.

\subsection{Light sfermion scenarios}\label{sec:Light_sfermions}
In this subsection we revisit the scenarios considered above, but allowing for a contribution to the anapole moment from fermion-sfermion loops.

\subsubsection*{Bino limit}

As discussed in subsection \ref{sec:pure_heavy_sfermion}, in the limit of heavy sfermions, the lightest neutralino does not couple to the $W$ boson nor to the charged Higgs, therefore ${\cal A}_W, {\cal A}_H\simeq 0$. On the other hand, the bino  couples to the Standard Model fermions and sfermions, and the sfermions in the loop could contribute sizably to the anapole moment if they are sufficiently light. We consider here a simplified scenario where the bino couples to the left- and right-handed components of a Standard Model fermion $f$ (with color charge $N_c$, electric charge $Q_f$ and isospin $T^f_{3L}$), and the sfermions $\tilde f_L$ and $\tilde f_R$. We denote the two scalar mass eigenstates as $\tilde f_1$ and $\tilde f_2$, which are obtained from the interaction eigenstates $\tilde f_L $ and $\tilde f_R$ by rotating by the angle $\theta_{\tilde f}$ (see Appendix \ref{sec:ParticleSpectrumMSSM}). The strength of the Yukawa coupling of the lightest neutralino to the sfermion mass eigenstates $\tilde f_1$ and $\tilde f_2$ and the left- and right-handed components of the Standard Model fermion $f$ explicitly read:
\begin{alignat}{2}
c^{1}_L&=-\sqrt{2}g\left[\tan\theta_W(Q_{f}-T^{f}_{3L})\right]\cos\theta_{\widetilde{f}} ,~~~
&&c^{1}_R=\sqrt{2}g\tan\theta_WQ_{f}\sin\theta_{\widetilde{f}}, \nonumber \\
c^{2}_L&=\sqrt{2}g\left[\tan\theta_W(Q_{f}-T^{f_i}_{3L})\right]\sin\theta_{\widetilde{f}
},~~~
&&c^{2}_R=\sqrt{2}g\tan\theta_WQ_{f}\cos\theta_{\widetilde{f}},
\end{alignat}
which are in general parity violating and therefore will generate a non-vanishing contribution to the anapole moment. 

We show in the top left panel of \cref{fig:ScatterPlotsLimits} a scatter plot of the expected anapole moment (normalized to the nuclear magneton) for the pure bino scenario for $m_\chi\in [10^1,10^4]$ GeV, $m_{\widetilde f_1} \in [m_\chi, 10 m_\chi]$, $m_{\widetilde f_2}\in [m_{\widetilde f_1}, 10 m_\chi]$, and $\theta_{\widetilde f}\in  [0,2\pi]$. In the plot we have taken for concreteness $m_f=m_\tau=1.7$ GeV, and we have imposed the constraints on the stau mass from ATLAS~\cite{ATLAS:2019gti} and from the LEP experiments \cite{Berggren:2001kb}. Generically, one finds $|{\cal A}|/\mu_{N}\sim 10^{-8} (m_\chi/100\,{\rm GeV})^{-2}$ fm, although there are a few points with $10^{-6}\,{\rm fm}\lesssim |\mathcal{A}|/\mu_N\lesssim 10^{-5}$ fm for $\mDm \lesssim 100\gev$ where the anapole moment is enhanced, corresponding to a compressed spectrum scenario where the stau mass is close to the bino mass. 

\subsubsection*{Higgsino limit}

In order to generate an anapole moment in this simplified scenario it is also necessary to introduce new light degrees of freedom with parity violating couplings. As for the bino limit analyzed above, we consider the scenario where the Higgsino couples to the left- and right-handed components of a Standard Model fermion $f$ and the sfermions $\widetilde f_L$ and $\widetilde f_R$, with mass eigenstates $\widetilde f_1$ and $\widetilde f_2$. The coupling strengths to the mass eigenstates explicitly read:
\begin{alignat}{2}
c^{1}_L&= H^{{f}_{L}} \sin\theta_{\widetilde{f}},\quad
&&c^{1}_R= H^{{f}_{L}} \cos\theta_{\widetilde{f}},\nonumber\\
c^{2}_L&= H^{{f}_{L}} \cos\theta_{\widetilde{f}},\quad
&&c^{2}_R=-H^{{f}_{L}} \sin\theta_{\widetilde{f}},
\end{alignat}
with
\begin{align}
H^{{f}_{L}} &=- \frac{g}{2m_W}m_{f} \times \begin{cases}
	1/\sin\beta, & f =u\text{-type}\\
	1/\cos\beta, & f = d\text{-type}, \ell\\
\end{cases}\,,
\end{align}
which are as before parity violating. 

We show in the top right panel of \cref{fig:ScatterPlotsLimits} a scatter plot of the anapole moment for the pure higgsino scenario, for the same range of parameters as for the pure bino scenario, and taking $\tan\beta=5$ (red points) or $\tan\beta=50$ (blue points). Clearly the anapole moment increases with $\tan\beta$, as the Higgsino coupling to the tau-stau grows with $\cos^{-2}\beta$.

\subsubsection*{Wino limit} 

Similarly to the previous two scenarios, in the wino limit the anapole moment can only be generated by parity-violating interactions of the lightest neutralino with fermions and sfermions. The coupling strengths to the sfermion mass eigenstates $\widetilde f_1$ and $\widetilde f_2$ read in this limit:
\begin{alignat}{2}
c^{i,1}_L&=-\sqrt{2}g T^{f_i}_{3L}\cos\theta_{\widetilde{f}},\quad
&&c^{i,1}_R=0, \nonumber \\
c^{i,2}_L&=\sqrt{2}g T^{f_i}_{3L}\sin\theta_{\widetilde{f}}, \quad
&&c^{i,2}_R=0,
\end{alignat}
which are clearly parity violating. 

\begin{figure}[t!]
	\centering
	\includegraphics[width=0.49\textwidth, clip = true]{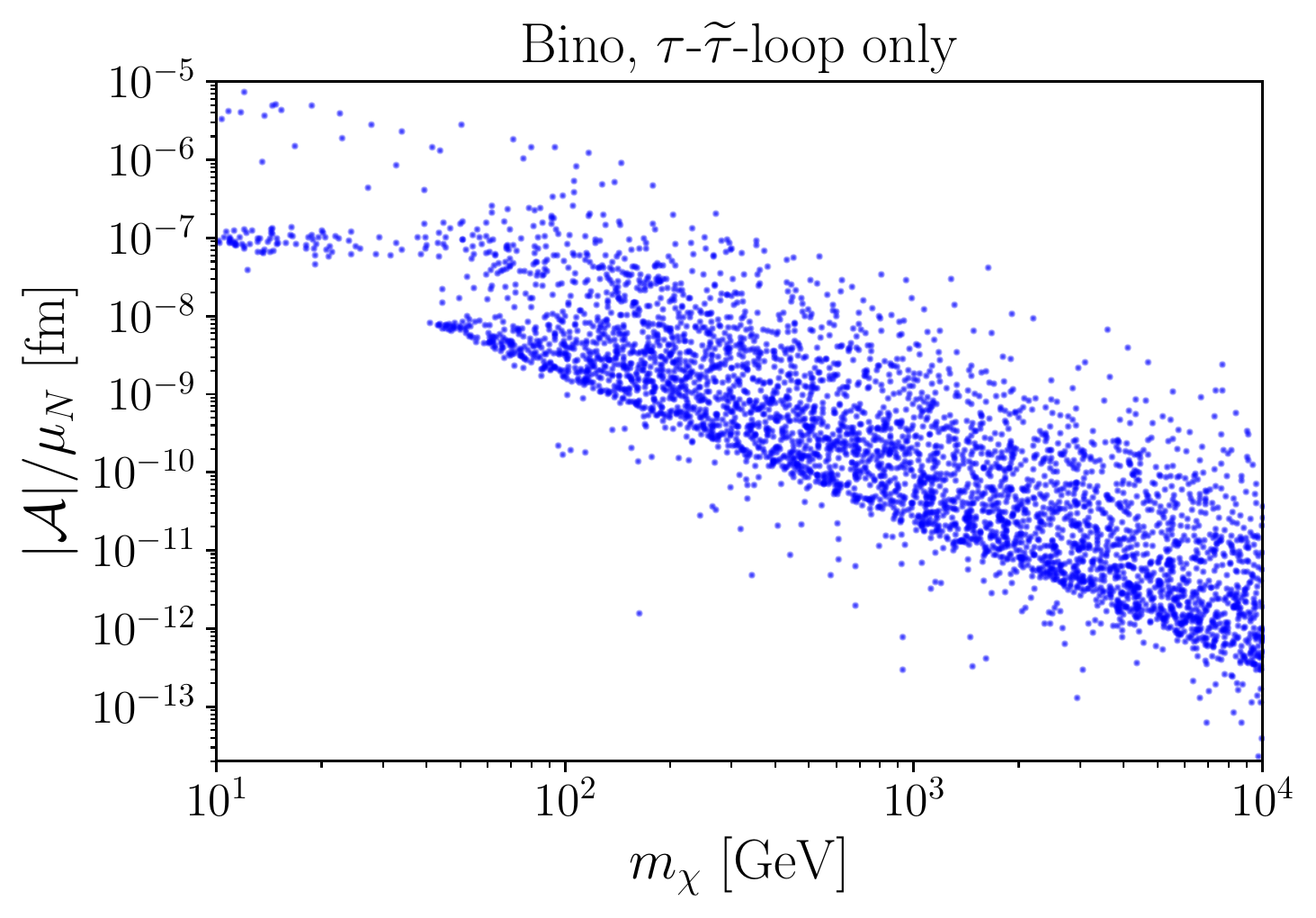}
	\includegraphics[width=0.49\textwidth, clip = true]{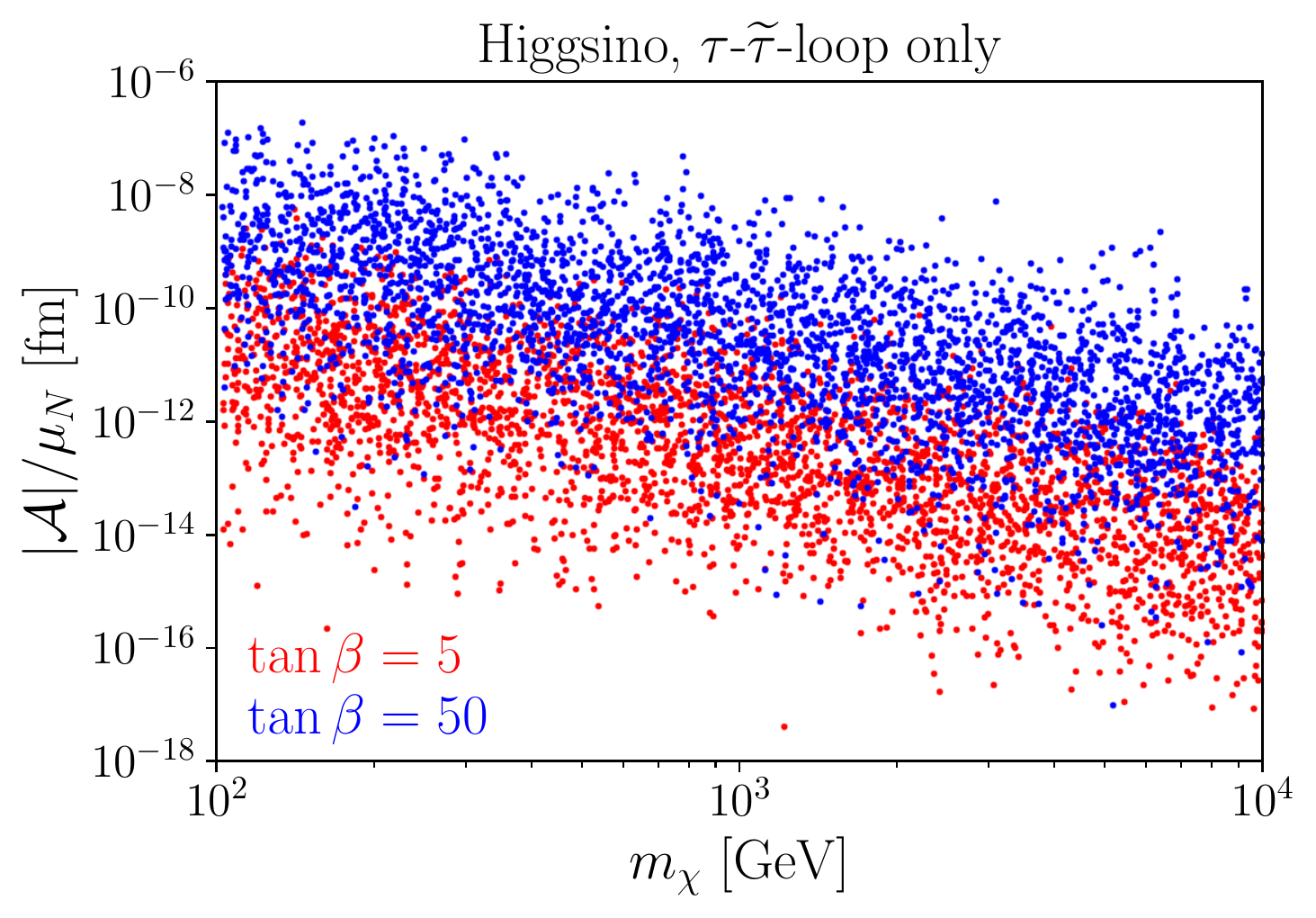}\\
	\includegraphics[width=0.49\textwidth,clip = true]{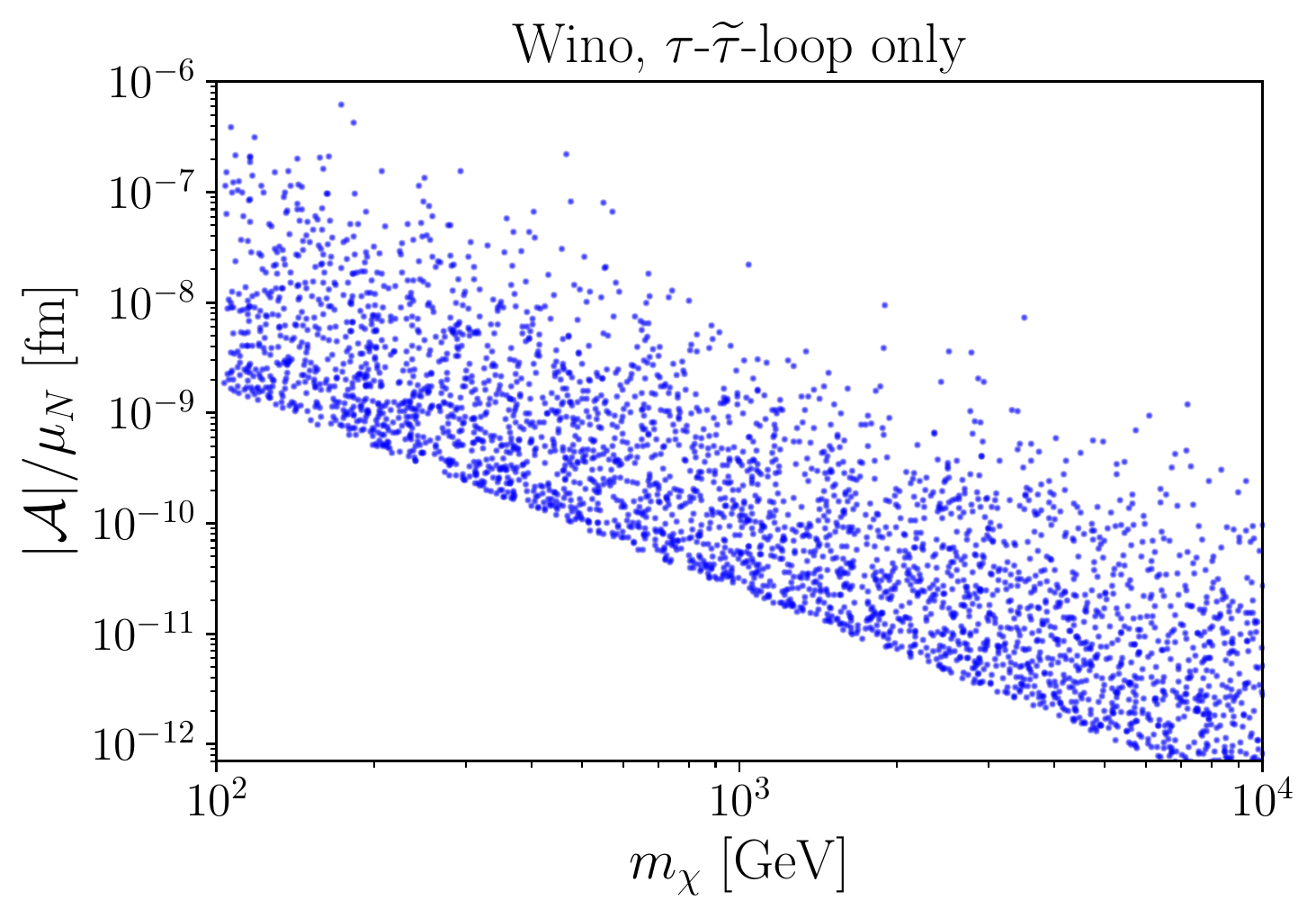}
	\caption{Anapole moment (normalized to the nucleon magneton) for simplified MSSM scenarios with pure bino (top left), pure higgsino (top right), and pure wino (bottom)  lightest neutralino, coupling only with the tau and staus. For details, see subsection \ref{sec:Light_sfermions}.}
	\label{fig:ScatterPlotsLimits}
\end{figure}

The expected anapole moment in this scenario is shown in the lower panel of \cref{fig:ScatterPlotsLimits}, for the same ranges of parameters as for the Bino limit.

\subsection{Mixed lightest neutralino \& decoupled sfermions}
\label{sec:mixed_heavy_sfermion}

Finally, we consider a simplified scenario where the sfermions are very heavy, so that ${\cal A}_{\widetilde f}\simeq 0$, but with an admixture of interaction eigenstates in the lightest neutralino mass eigenstates, which may allow parity violating interactions. 

\subsubsection*{Mixed Bino-Higgsino}

In the limit $M_1,\mu \ll M_2$, there is only one light chargino, which is purely a charged Higgsino. The couplings of the lightest neutralino to the charged Higgsino and the $W$ are:
\begin{alignat}{2}
&v^1_L =  - g\frac{1}{\sqrt2} N_{13},\quad &&v^1_R =   +g\frac{1}{\sqrt2} N_{14}^*, \nonumber \\
&c_L^{G,1} =-\frac{g}{\sqrt{2}}\cos\beta\tan\theta_W N_{11}^*, \quad
&&c_R^{G,1} =-\frac{g}{\sqrt{2}}\sin\beta\tan\theta_W N_{11},
\end{alignat}
which are in general parity violating, thus leading to a non-zero ${\cal A}_W$. The couplings to the chargino and the charged Higgs read
\begin{align}
		c_L^{H,1} = g\sin\beta\tan\theta_W N_{11}^*, ~~~~~
		c_R^{H,1} =-g\cos\beta\tan\theta_W N_{11},
\end{align}
which are also in general parity violating and can further increase the anapole moment. 
 
We show in the left panel in \cref{fig:ScatterPlotsLimitsMixed} the expected anapole moments for this scenario, taking for concreteness  $M_1$, $\mu\in [100,10^5]\gev$ and $\tan\beta=5$. We assume for simplicity that the charged Higgs is very heavy and does not contribute to the anapole moment (although clearly for a light charged Higgs the anapole moment could be enhanced). In the plot we also indicate whether the lightest neutralino  is bino like ($|N_{11}|>0.95$), higgsino like ($\sqrt{N_{13}^2+N_{14}^2}>0.95$) or a mixed state. As expected, the anapole moment is enhanced when the lightest neutralino is not a pure state, but an admixture of bino and higgsino. 

\subsubsection*{Mixed Bino-Wino} 

In the limit $M_1, M_2\ll \mu$, there is only one light chargino, which is purely a charged wino. The couplings of the lightest neutralino to the charged Higgsino and the W are:
\begin{alignat}{2}\label{eq:MSSMVectorCouplings_WB}
&v^1_L = -gN_{12}, \quad
&&v^1_R = -gN_{12}^*,   \nonumber \\
&c_L^{G,1} =0, \quad
&&c_R^{G,1} =0,
\end{alignat}
which preserve parity and therefore give ${\cal A}_W\simeq 0$. Further, the couplings to the chargino and the charged Higgs are:
\begin{align}
		c_L^{H,1} = 0,\quad
		c_R^{H,1} =0.
\end{align}
Therefore, also in the scenario where the lightest neutralino is a mixed bino-wino  state, only the sfermion loops can generate a non-vanishing anapole moment.  

\subsubsection*{Mixed Wino-Higgsino}

\begin{figure}[t!]
	\centering
		\includegraphics[width=0.49\textwidth, clip = true]{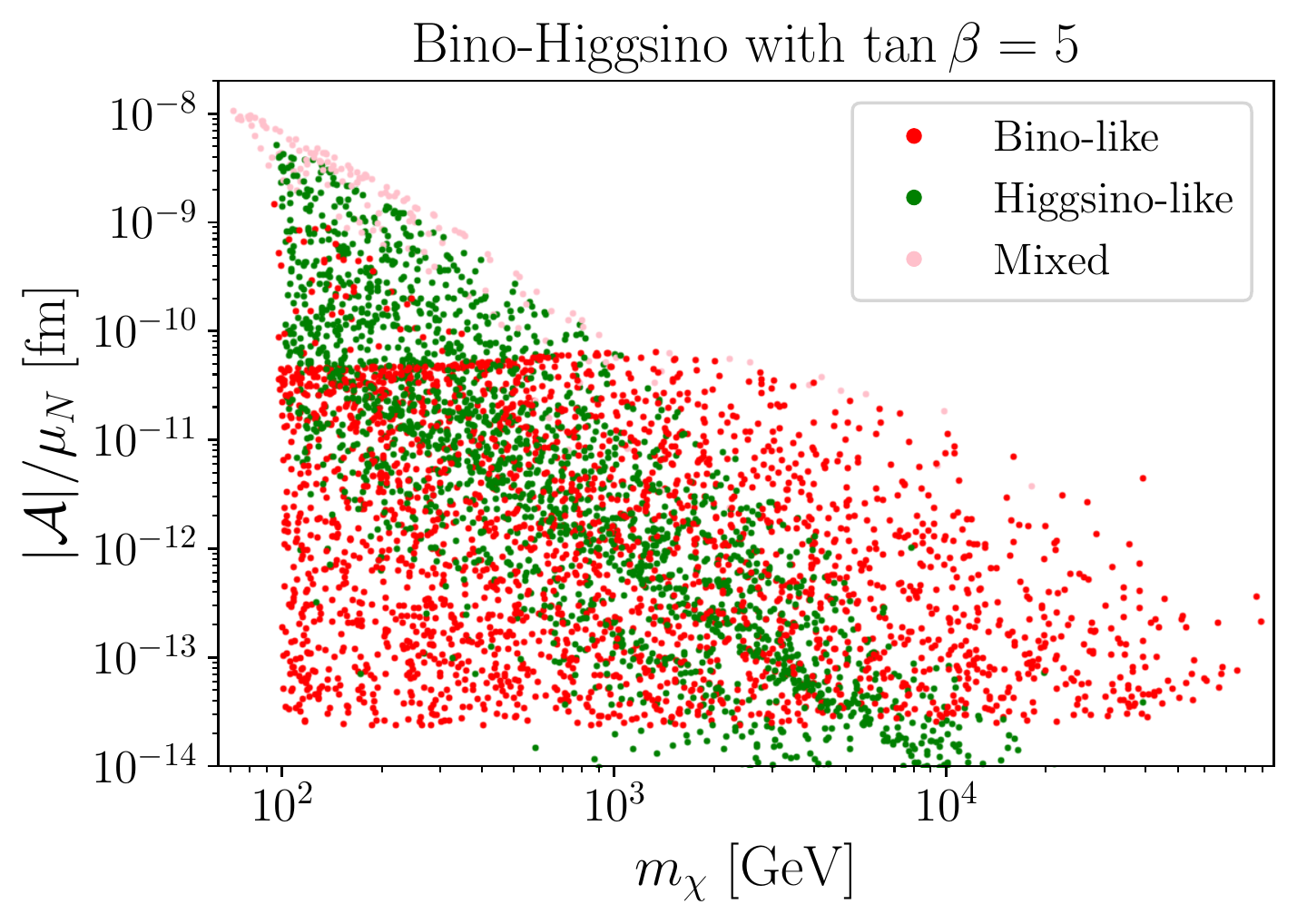}
		\includegraphics[width=0.49\textwidth,clip= true]{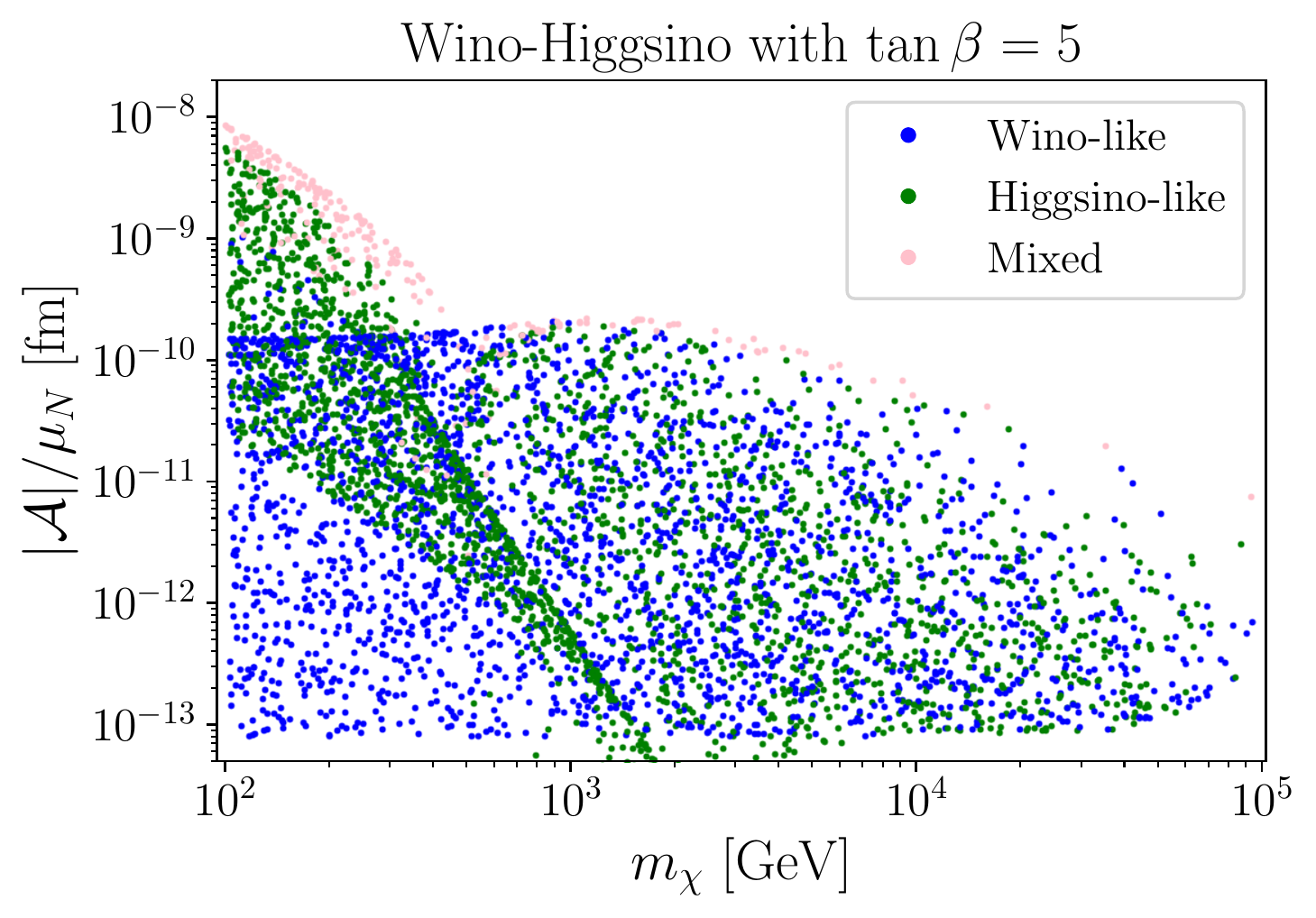}
	\caption{Anapole moment (normalized to the nucleon magneton) for simplified MSSM scenarios with mixed bino-higgsino (left) and wino-higgsino (right) lightest neutralino For details, see subsection \ref{sec:mixed_heavy_sfermion}.}
	\label{fig:ScatterPlotsLimitsMixed}
\end{figure}

In the limit $M_2,\mu\ll M_1$ both charginos can be light and contribute to the anapole moment via the interactions with the $W$ and with the charged Higgs boson. The coupling strengths of the lightest neutralino to the  charginos $\chi^\pm_i$, $i=1,2$ and the $W$ boson read:
\begin{align}\label{eq:MSSMVectorCouplings_WH}
v^j_L &= -gN_{12} U_{j1}^* - g\frac{1}{\sqrt2} N_{13} U_{j2}^*, \nonumber \\
v^j_R &= -gN_{12}^* V_{j1}   +g\frac{1}{\sqrt2} N_{14}^* V_{j2}, \nonumber\\
c_L^{G,j} &=g\cos\beta\left[N_{13}^* U_{j1}^*-\frac{1}{\sqrt 2}U_{j2}^*(N_{12}^*)\right], \nonumber\\
c_R^{G,j} &=-g\sin\beta\left[N_{14}V_{j1}+\frac{1}{\sqrt 2} V_{j2}(N_{12})\right],
\end{align}
while for the charged Higgs boson,
\begin{align}
		c_L^{H,j} &= -g\sin\beta\left[N_{13}^* U_{j1}^*-\frac{1}{\sqrt 2}U_{j2}^*(N_{12}^*)\right], \nonumber \\
		c_R^{H,j} &=-g\cos\beta\left[N_{14}V_{j1}+\frac{1}{\sqrt 2} V_{j2}(N_{12})\right].
\end{align}
These interactions are in general parity violating and lead to a non-vanishing anapole moment. A numerical scan of this scenario is shown in the right panel in \cref{fig:ScatterPlotsLimitsMixed}, for the same set-up as in the bino-higgino mixed case; the conclusions in the wino-higgsino case are also analogous to that case.

\subsection{General MSSM scenarios}
\label{sec:FullMSSM}

So far we have concentrated in some limiting scenarios where most SUSY particles are assumed to be very heavy and integrated-out. On the other hand, in generic scenarios several SUSY particles can be light and can contribute sizeable to the anapole moment. To estimate the anapole moment expected in a generic SUSY scenario, we will consider in this subsection MSSM scenarios defined by the ranges  indicated in table \ref{table:ParameterSpaceFullMSSM} (at the scale $\Lambda=3$ TeV).

\begin{table}
	\centering
\begin{tabular}{|c|c|}
	\hline
Parameter	& Range    \\ \hline
 $M_1$ &  $[100, 2000]$ GeV  \\ \hline
$M_2$ &  $[100, 2000]$  GeV \\ \hline
$M_3$ &  $[2000,5000]$ GeV \\ \hline
$A_{t,b,\tau}$	& $[-4000,4000]$ GeV   \\ \hline
$m_A$	& $[10^3, 10^5]$ GeV   \\ \hline
$\tan\beta$	& $[3,50]$  \\ \hline
$\mu$	& $[100,2000]$ GeV \\ \hline
$m_{\widetilde{\ell}_{L,R}}$ & $[100, 2000]$ GeV   \\ \hline
$m_{\widetilde{q}_{L_{1,2}}}$	& $[400,2000]$ GeV  \\ \hline
$m_{\widetilde{u}_{R_{1,2}}}$  , $m_{\widetilde{d}_{R_{1,2}}}$	& $[400, 2000]$ GeV   \\ \hline
$m_{\widetilde{q}_{L_3}}$	& $[300,2000]$ GeV  \\ \hline
$m_{\widetilde{u}_{R_3}}$, $m_{\widetilde{d}_{R_3}}$	& $[300,2000]$ GeV   \\ \hline
\end{tabular}
\caption{Ranges of parameters, defined at the scale $\Lambda=3$ TeV, for the MSSM scan described in subsection \ref{sec:FullMSSM}.}
\label{table:ParameterSpaceFullMSSM}
\end{table}

\begin{figure}[t!]
	\centering
	\includegraphics[width=.7\linewidth]{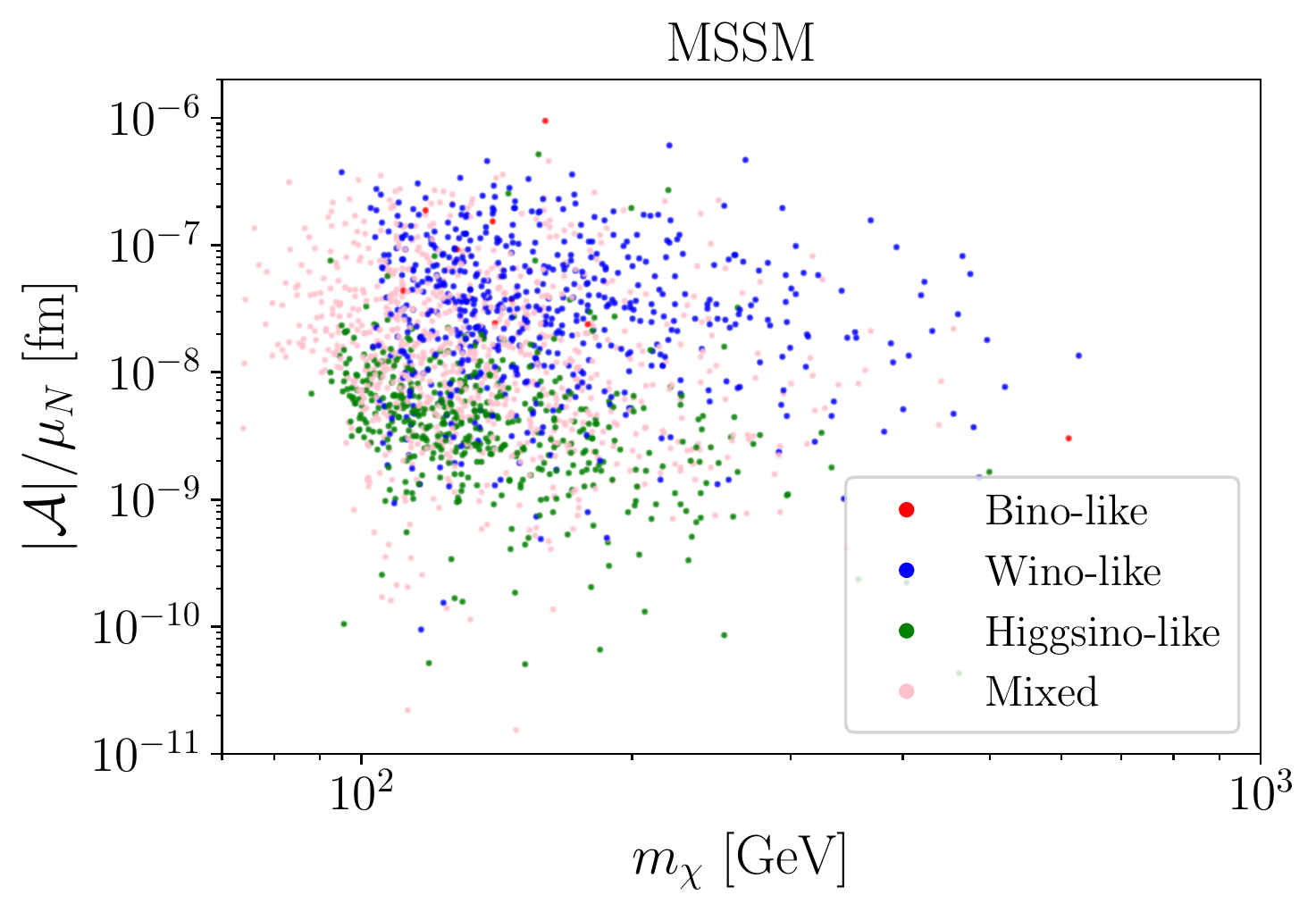}
	\caption{Anapole moment for MSSM scenarios with parameters in the ranges listed in Table  \ref{table:ParameterSpaceFullMSSM}  and satisfying the experimental constraints discussed in subsection \ref{sec:FullMSSM}.}
	\label{fig:pMSSM_without_DD}
\end{figure}

From those boundary conditions, we generate the low energy spectrum using \texttt{SOFTSUSY4.0} \cite{Allanach:2017hcf}. We then select the points satisfying the LEP constraints (using  \texttt{micrOMEGAs v3} \cite{Belanger:2013oya}), and ATLAS and CMS constraints (using \texttt{SModelS v2} \cite{Alguero:2021dig}), leading to a Higgs boson with mass in the range 123-127 GeV (using \texttt{HiggsBounds v4} \cite{Bechtle:2013wla} and \texttt{HiggsSignals} \cite{Stal:2013hwa}), and satisfying various flavor physics constraints (using  \texttt{SuperIso v3.0} \cite{Mahmoudi:2009zz} and \texttt{GM2Calc} \cite{Athron:2015rva}).~\footnote{We used \texttt{PySLHA} \cite{Buckley:2013jua} for linking the various codes via the SLHA \cite{Allanach:2008qq} format.}.  These points do not necessarily reproduce the observed dark matter abundance in the standard freeze-out mechanism, although they could become viable for other production mechanisms. Since we are interested in the generic size of the neutralino anapole moment, we will disregard in our analysis the constraints from Cosmology. The resulting values of the anapole moment, calculated using the general expressions from  \cref{sec:AnapoleMoment_MSSM}, are shown in  \cref{fig:pMSSM_without_DD}.

We find points where the anapole moment can reach values up to $|{\cal A}|/\mu_N\sim 10^{-6}\,\fm$. These correspond to scenarios where the anapole moment is dominated by the fermion-sfermion contribution and where the LSP and a sfermion are almost mass-degenerate, in accordance with the results for the simplified models of \cref{sec:Light_sfermions} (with ${\cal O}(1)$ enhancements when several sfermions circulate in the loop), and correspond to scenarios where the lightest neutralino contains a significant bino and/or wino component. For scenarios where the lightest neutralino is Higgsino like, the anapole moment is typically more suppressed.

\section{Direct Dark Matter Detection Through the Anapole Moment}\label{sec:DDlimits_pMSSM}

The effective Lagrangian \cref{eq:EffectiveLagrangianAnapole} gives rise to a dark matter interaction with the nuclei, that can induce an observable signal in direct detection experiments.\footnote{More strictly, the contact interaction approximation holds when the momentum transfer is smaller than the masses in the loop, which we assume here. For coupling to electrons, a momentum-dependent form factor should instead be considered \cite{Kopp:2014tsa}.}
The differential scattering cross section induced by the interaction of the Majorana dark matter particle with a target nucleus via the anapole moment reads \cite{Ho:2012bg,DelNobile:2014eta}:
\begin{align}\label{directdetection:eq:AnapoleDiffxsec}
\frac{d\sigma}{dE_R}
=
\alpha_{\rm{EM}}\,
\mathcal{A}^2
\biggl[
Z^2\biggl(2m_T-\biggl(1+\frac{m_T}{m_\chi}\biggr)^2\frac{E_R}{v^2}\biggr)F^2_Z(q^2)
+
\frac{1}{3}\frac{m_T}{m^2_\chi}\l(\frac{\bar{\mu}_T}{\mu_N}\r)^2\frac{E_R}{v^2}F^2_D(q^2)
\biggr]
\,,
\end{align}
where $m_T$ and $Z$ are the nucleus mass and electric charge, $E_R$ is the recoil energy (related to the  momentum transfer through $q^2=2 m_T E_R$) and $v$ is the dark matter speed relative to the nucleus. Further,
$F_Z$ and $F_D$ are the  charge and magnetic dipole moment form factors~\cite{Helm:1956zz,Lewin:1995rx}: 
\begin{align}
F^2_Z(q^2)
&=
\l(\frac{3j_1(qR)}{qR}\r)^2 e^{-q^2s^2}, \\
F^2_D(q^2)
&=
\left\{
\begin{array}{ll}
\l[\frac{\sin(qR_D)}{qR_D}\r]^2 & (qR_D<2.55, qR_D>4.5)\\
0.047 & (2.55\leq qR_D\leq4.5)
\end{array}
\right.\,.
\end{align}
where $j_1(x)$ is a spherical Bessel function of the first kind, $R=\sqrt{c^2+\frac{7}{3}\pi^2 a^2-5s^2}$ (with $c=(1.23A^{1/3}-0.60)\,\mbox{fm}$, $a=0.52\,\mbox{fm}$ and $s=0.9\,\mbox{fm}$) and $R_D\simeq 1.0A^{1/3}$ fm. $A$ denotes the mass number of target nuclei. Further,  ${\mu}_N=e/2m_p$ denotes the nuclear magneton,  and $\bar \mu_T$ is the weighted dipole moment for the target nuclei, defined as:
\begin{align}
\bar{\mu}_T
=
\l(
\sum_{i}f_i\mu_{i}^2\frac{S_{i}+1}{S_{i}}
\r)^{1/2}\,,
\end{align} 
where $f_i$, $\mu_{i}$, and $S_{i}$ are the elemental abundance, nuclear magnetic moment, and spin, respectively, of the isotope $i$ \cite{Chang:2010en}.

The differential event rate at a direct detection experiment reads:
\begin{align}
\frac{dR}{dE_R}
=
\frac{1}{m_T}\frac{\rho_{\rm{loc}}}{m_\chi}\int d^3v\, v f_{\rm{Lab}}({\vec{v}})\frac{d\sigma}{dE_R}\;,
\label{eq:dRdE}
\end{align}
where $\rho_{\rm{loc}}=0.3\,\mbox{GeV}\,\mbox{cm}^{-3}$ and $f_{\rm{Lab}}(\vec{v})$ denotes the dark matter velocity distribution in the laboratory frame. For the latter, we will adopt a Maxwell-Boltzmann distribution in the galactic frame, truncated at the escape velocity from the Galaxy, $v_{\text{esc}}$:
\begin{equation}
f_{\rm{Lab}}(\vec{v}) = f (\vec{v} + \vec{v}_\text{E}) ~,
\end{equation}
with $\vec{v}_{\text{E}}$ the velocity of the Earth in the galactic frame and 
\begin{equation}
f (\vec{v}) = 
\begin{cases}
\frac{1}{\cal N} e^{-v^2/v_0^2}  & (|\vec{v}| < v_{\text{esc}}) \\ 
0& (|\vec{v}| > v_{\text{esc}}) 
\end{cases}
~,
\end{equation}
with 
\begin{equation}
{\cal N} = \pi^{3/2} v_0^3 \biggl[
\text{erf} \biggl(\frac{v_{\text{esc}}}{v_0}\biggr)
- \frac{2 v_{\text{esc}}}{\sqrt{\pi} v_0}  e^{-
	\frac{v_{\text{esc}}^2}{v_0^2}} 
\biggr]~.
\end{equation}
Hereafter we take 
$v_{\rm{esc}}=544\,{\rm km}\,{\rm s}^{-1}$,
$v_{0}=220\,{\rm km}\,{\rm s}^{-1}$
and
$v_{\rm E}=232\,{\rm km}\,{\rm s}^{-1}$. 
Finally, we calculate the number of events at a given direct detection experiment integrating $dR/dE_R$ over the recoil energy, taking into account the corresponding detection efficiency.

\begin{figure}[t!]
	\centering
	\includegraphics[width=.6\linewidth]{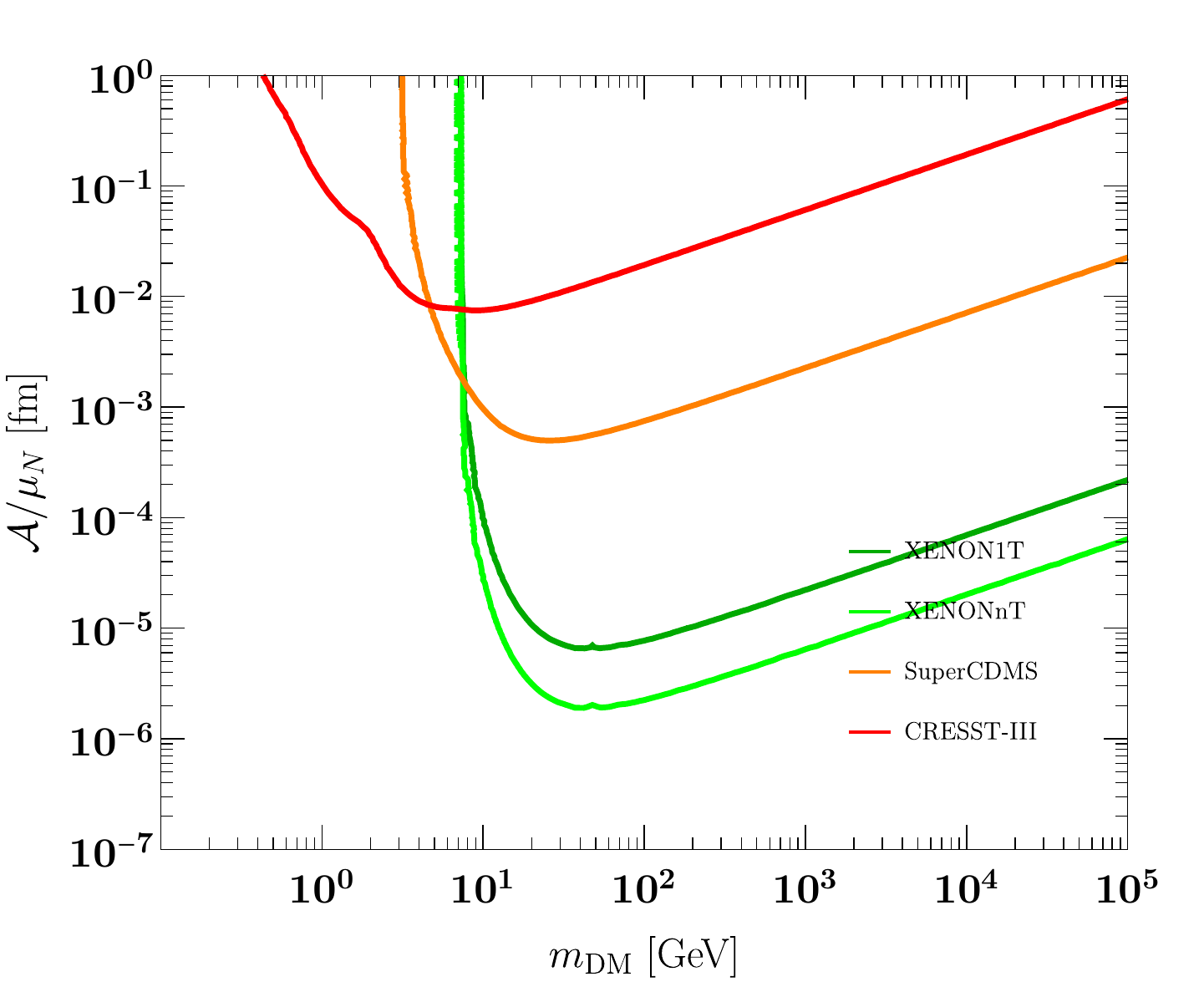}
	\caption{Upper limit on the anapole moment of a Majorana spin 1/2 fermion as dark matter candidate from the XENON1T, SuperCDMS and CRESST-III data, and projected sensitivity for XENONnT.}
	\label{fig:pMSSM_with_DD}
\end{figure}
We show in \cref{fig:pMSSM_with_DD} the 90\% C.L. upper limits on the anapole moment $\mathcal{A}$ normalized by the nuclear magneton $\mu_N$ from the non-observation of a dark matter signal at the XENON1T \cite{Aprile:2018dbl}, SuperCDMS \cite{Agnese:2014aze}, and CRESST-III \cite{Amole:2019fdf} experiments, alongside with the expected sensitivity of the XENONnT experiment \cite{Aprile:2015uzo}.~\footnote{ Details of the estimation of the detection efficiency are given in Appendix B of Ref.~\cite{Hisano:2020qkq}.} We find that the current sensitivity from the XENON experiment reaches ${\cal A}/\mu_N \sim 10^{-5}\,\fm$ at $m_{\rm DM}\sim 30$ GeV, which is about one order of magnitude larger than the maximum anapole moment we predict for generic MSSM scenarios. For these scenarios, it would be necessary to improve in sensitivity by at least one order of magnitude in order to probe the anapole moment of a spin 1/2 Majorana dark matter candidate, unless the Earth is immersed in a region of the galaxy with an overdensity of dark matter. Let us note that for special choices of parameters, namely when the dark matter candidate is almost degenerate in mass with the scalar (or vector) in the loop and when the fermion is very light, the anapole moment is enhanced ({\it cf.} Figs.~ \ref{fig:VectorAnapolePlot} and \ref{fig:ScalarAnapolePlot}, and also \cite{Garny:2015wea}). In these very special cases, a signal might be expected. 

In this section we have considered simplified scenarios where the dark matter only interacts with the nucleon via the anapole moment. Clearly, there could be MSSM scenarios where the scattering mediated by squarks or by Higgses dominate over the one mediated by the anapole moment. For those scenarios, the discovery potential of dark matter accordingly increases. However, establishing the existence of an electromagnetic multiple moment for Majorana dark matter will become even more challenging.

%%%%%%%%%%%%%%%%%%%%
\section{Conclusions}\label{sec:conclusions}

In this work we have calculated the leading contribution to the anapole moment of a spin 1/2 Majorana fermion that interacts via a Yukawa or a gauge interaction with electromagentically charged particles. To ensure the finiteness and the gauge independence of the vector contribution, we employed the background field method. 

We  have applied our general results to calculate the anapole moment of the lightest neutralino in the Minimal Supersymmetric Standard Model. Since the anapole interaction violates parity, only those MSSM scenarios violating parity will generate a non-vanishing anapole moment. We have also studied various limits where many supersymmetric particles are integrated out, and which can be identified with simplified dark matter models where the dark matter candidate is a Majorana fermion, that transforms as a singlet, doublet or triplet of $SU(2)_L$, and which could interact with a fermion and a sfermion via a Yukawa coupling. 

Lastly, we have derived upper limits on the anapole moment of a Majorana fermion as dark matter candidate, from the null search results of direct detection experiments. For the parameters of the Standard Halo Model, we find that an improvement of sensitivity of current experiments by at least one order of magnitude would be necessary in order to probe the anapole moment of generic dark matter scenarios.

\acknowledgments 
The work of A.I. and M.R. was supported by the Collaborative Research Center SFB1258 and by the Deutsche Forschungsgemeinschaft (DFG, German Research Foundation) under Germany's Excellence
Strategy - EXC-2094 - 390783311. 
The Feynman diagrams in this paper were drawn with the help of \texttt{TikZ-Feynman} \cite{ELLIS2017103}.

\appendix
\section{Particle Spectrum of the MSSM} \label{sec:ParticleSpectrumMSSM}

In this Appendix we summarize the calculation of the particle mass eigenstates in terms of the interaction eigenstates in the Minimal Supersymmetric Standard Model (for reviews, see {\it e.g.} \cite{
Haber:1984rc,Drees:2004SUSY}).

The R-parity odd neutral fermions of the MSSM are the bino ($\widetilde B$), the neutral wino ($\widetilde W^0$) and the two neutral Higgsinos ($\widetilde{H}^0_1$ and $ \widetilde{H}^0_2$), with mass terms that can be cast as:
\begin{align}
-\mathcal{L}
=
\frac{1}{2}
\l(
\begin{array}{cccc}
\widetilde B
&
\widetilde W^0
&
\widetilde{H}^0_1
&
\widetilde{H}_0^2
\end{array}
\r)
\mathcal{M}^n
\l(
\begin{array}{c}
\widetilde B
\\
\widetilde W^0
\\
\widetilde{H}^0_1
\\
\widetilde{H}^0_2
\end{array}
\r)
+h.c.,
\end{align}
where  $\mathcal{M}^n$ is given by
\begin{align}
\mathcal{M}^n
=
\l(
\begin{array}{cccc}
M_1 & 0 & -m_Z\sin\theta_W\cos\beta & m_Z\sin\theta_W\sin\beta \\
0 & M_2 & m_Z\cos\theta_W\cos\beta & -m_Z\cos\theta_W\sin\beta \\
-m_Z\sin\theta_W\cos\beta & m_Z\cos\theta_W\cos\beta & 0 & -\mu\\
m_Z\sin\theta_W\sin\beta & -m_Z\cos\theta_W\sin\beta & -\mu & 0
\end{array}
\r)\,.
\end{align}
Here, $M_1$ and $M_2$ are respectively the bino and wino masses, $\mu$ is the Higgsino mass parameter,  $\theta_W$ is the weak mixing angle, and $\tan\beta\equiv \langle H_2^0\rangle/\langle H_1^0\rangle$ is the ratio of Higgs vacuum expectation values. The neutralinos $\chi_i$, $i=1,2,3,4$ are defined as the R-parity-odd neutral fermion mass eigenstates, and are constructed by diagonaling the mass matrix ${\cal M}^n$. To this end, one introduces the unitary matrix  $N$, defined such that
\begin{align}
N^*\mathcal{M}^n N^\dag =
\mbox{diag}\l(m_{\chi_1}, m_{\chi_2}, m_{\chi_3}, m_{\chi_4}\r)\,,
\end{align}
with $m_{\chi_i}$ the neutralino masses, defined as real and positive, and ordered so that $m_{\chi_1}\leq m_{\chi_2}\leq m_{\chi_3} \leq m_{\chi_4}$. The neutralino states are related to the interaction eigenstates through:
\begin{align}
\l(
\begin{array}{ccc}
\chi_1\\
\chi_2\\
\chi_3\\
\chi_4
\end{array}
\r)
=
N
\l(
\begin{array}{c}
\widetilde B\\
\widetilde W^0\\
\widetilde{H}^0_1\\
\widetilde{H}^0_2\\
\end{array}
\r)
\,.
\end{align}

Similarly, the R-parity odd charged fermions of the MSSM are the charged winos ($\widetilde W^\pm$) and the two charged Higgsinos ($\widetilde{H}^-_1$ and $ \widetilde{H}^+_2$). Their mass terms have the form: 
\begin{align}
-\mathcal{L}
=
\l(
\begin{array}{cc}
\widetilde W^+
&
\widetilde{H}^+_2
\end{array}
\r)
\mathcal{M}^c
\l(
\begin{array}{c}
\widetilde W^-
\\
\widetilde{H}^-_1
\end{array}
\r)
+h.c.,
\end{align}
where
\begin{align}
\mathcal{M}^c
=
\l(
\begin{array}{cc}
M_2 & \sqrt{2}m_W\cos\beta \\
\sqrt{2}m_W\sin\beta & \mu
\end{array}
\r)\,.
\label{eq:MC}
\end{align}
with $m_W$ the W-boson mass. The charginos $\chi_i^\pm$, $i=1,2$ are defined as the R-parity odd charged fermion mass eigenstantes, and are constructed from diagonalizing the mass matrix as
\begin{align}
V\mathcal{M}^c U^T
=
\mbox{diag}\l(m_{\chi^-_1},m_{\chi^-_2}\r)
\,,
\end{align}
with $m_{\chi^-_1}\leq m_{\chi^-_2}$ and $m_{\chi^-_i}$ being real and positive. The charginos are related to the interaction eigenstates through:
\begin{align}
\l(
\begin{array}{ccc}
\chi^-_{1}\\
\chi^-_{2}
\end{array}
\r)
=
U
\l(
\begin{array}{ccc}
\widetilde W^-
\\
\widetilde{H}^-_1
\end{array}
\r)\,,
\qquad
\l(
\begin{array}{ccc}
\chi^+_{1}\\
\chi^+_{2}
\end{array}
\r)
=
V
\l(
\begin{array}{ccc}
\widetilde W^+
\\
\widetilde{H}^+_2
\end{array}
\r)\,,
\end{align}
where $U$ and $V$ can be parameterized as
\begin{align}
U
=
\l(
\begin{array}{cc}
\cos\phi_L & \sin\phi_L \\
-\sin\phi_L & \cos\phi_L
\end{array}
\r)
\,,~~~
V
=
\l(
\begin{array}{cc}
\cos\phi_R & \sin\phi_R \\
-\epsilon_R\sin\phi_R & \epsilon_R\cos\phi_R
\end{array}
\r)
\,.
\end{align}
with
\begin{align}
&\tan2\phi_L
=
\frac{2\sqrt{2}m_W(\mu \sin\beta+M_2\cos\beta)}
{M^2_2-\mu^2-2m^2_W\cos2\beta}
\,,\\
&\tan2\phi_R
=
\frac{2\sqrt{2}m_W(\mu \cos\beta+M_2\sin\beta)}{M^2_2-\mu^2+2m^2_W\cos2\beta}
\,,\\
&\epsilon_R
=
\mbox{sgn}\biggl(M_2 \mu -m^2_W\sin2\beta
\biggr)
\,.
\end{align}

Finally, we focus on the sfermion mass term.
The mass matrix for the superpartners of the SM fermion $f$ reads:
\begin{align}
-\mathcal{L}
=
\l(
\begin{array}{cc}
\widetilde{f}^*_L
&
\widetilde{f}^*_R
\end{array}
\r)
\widetilde{\mathcal{M}}^2_f
\l(
\begin{array}{c}
\widetilde{f}_L
\\
\widetilde{f}_R
\end{array}
\r)
+h.c.,
\end{align}
where
\begin{align}
\widetilde{\mathcal{M}}^2_f
=
\l(
\begin{array}{cc}
\mathcal{M}^2_{f_{LL}}
&
\mathcal{M}^2_{f_{LR}}\\
(\mathcal{M}^2_{f_{LR}})^*
&
\mathcal{M}^2_{f_{RR}}
\end{array}
\r)\,,
\end{align}
with
\begin{align}
&\mathcal{M}^2_{f_{LL}}
=
m^2_{\widetilde{f}_L}
+
m^2_f+m^2_Z\cos2\beta(T_{3f}-Q_f\sin^2\theta_W)
\,,\\
&\mathcal{M}^2_{f_{RR}}
=
m^2_{\widetilde{f}_R}
+
m^2_f+m^2_Z\cos2\beta Q_f\sin^2\theta_W
\,,\\
&\mathcal{M}^2_{f_{LR}}
=
\begin{dcases}
m_u(A_u-\mu \cot\beta)
& \mbox{for $f=u$ (up-type quark)}\\
m_f(A_f+\mu \tan\beta)
& \mbox{for $f=d,l$ (down-type quark, lepton)}
\end{dcases}
\,.
\end{align}
Here $T_{3f}$ and $Q_f$ are respectively the third component of isospin and the electric charge of the fermion $f$, 
$m^2_{\widetilde{f}_L}$ and 
$m^2_{\widetilde{f}_R}$ are soft SUSY breaking masses for the left- and right-handed chiral superfields, and $A_f$ are soft SUSY breaking trilinear terms. 
The mass matrix can be diagonalized as
\begin{align}
O_f\widetilde{\mathcal{M}}^2_f O^T_f
=
\mbox{diag}\l(m^2_{\widetilde{f}_1},m^2_{\widetilde{f}_2}\r)
\,,
\end{align}
with $m^2_{\widetilde{f}_1}\leq m^2_{\widetilde{f}_2}$ and $m^2_{\widetilde{f}_1}$ being real and positive. The sfermion mass eigenstates are related to the interaction eigenstates through:
\begin{align}
\l(
\begin{array}{ccc}
\widetilde{f}_1\\
\widetilde{f}_2
\end{array}
\r)
=\l(
\begin{array}{cc}
\cos\theta_f & \sin\theta_f\\
-\sin\theta_f & \cos\theta_f
\end{array}
\r)
\l(
\begin{array}{ccc}
\widetilde{f}_L\\
\widetilde{f}_R
\end{array}
\r)\,.
\end{align}

%%%%%%%%%%%%%%%%%%
\bibliographystyle{JHEP}
\bibliography{papers}

\providecommand{\href}[2]{#2}\begingroup\raggedright\begin{thebibliography}{10}

\bibitem{Jungman:1995df}
G.~Jungman, M.~Kamionkowski and K.~Griest, \emph{{Supersymmetric dark matter}},
  \href{https://doi.org/10.1016/0370-1573(95)00058-5}{\emph{Phys. Rept.}
  {\bfseries 267} (1996) 195}
  [\href{https://arxiv.org/abs/hep-ph/9506380}{{\ttfamily hep-ph/9506380}}].

\bibitem{Bertone:2004pz}
G.~Bertone, D.~Hooper and J.~Silk, \emph{{Particle dark matter: Evidence,
  candidates and constraints}},
  \href{https://doi.org/10.1016/j.physrep.2004.08.031}{\emph{Phys. Rept.}
  {\bfseries 405} (2005) 279}
  [\href{https://arxiv.org/abs/hep-ph/0404175}{{\ttfamily hep-ph/0404175}}].

\bibitem{Bergstrom:2000pn}
L.~Bergstr\"om, \emph{{Nonbaryonic dark matter: Observational evidence and
  detection methods}},
  \href{https://doi.org/10.1088/0034-4885/63/5/2r3}{\emph{Rept. Prog. Phys.}
  {\bfseries 63} (2000) 793}
  [\href{https://arxiv.org/abs/hep-ph/0002126}{{\ttfamily hep-ph/0002126}}].

\bibitem{Feng:2010gw}
J.~L. Feng, \emph{{Dark Matter Candidates from Particle Physics and Methods of
  Detection}},
  \href{https://doi.org/10.1146/annurev-astro-082708-101659}{\emph{Ann. Rev.
  Astron. Astrophys.} {\bfseries 48} (2010) 495}
  [\href{https://arxiv.org/abs/1003.0904}{{\ttfamily 1003.0904}}].

\bibitem{Fujikawa:1980yx}
K.~Fujikawa and R.~Shrock, \emph{{The Magnetic Moment of a Massive Neutrino and
  Neutrino Spin Rotation}},
  \href{https://doi.org/10.1103/PhysRevLett.45.963}{\emph{Phys. Rev. Lett.}
  {\bfseries 45} (1980) 963}.

\bibitem{Petcov:1976ff}
S.~T. Petcov, \emph{{The Processes $\mu \rightarrow e + \gamma, \mu \rightarrow
  e + \overline{e}, \nu' \rightarrow \nu + \gamma$ in the Weinberg-Salam Model
  with Neutrino Mixing}}, {\emph{Sov. J. Nucl. Phys.} {\bfseries 25} (1977)
  340}.

\bibitem{Pal:1981rm}
P.~B. Pal and L.~Wolfenstein, \emph{{Radiative Decays of Massive Neutrinos}},
  \href{https://doi.org/10.1103/PhysRevD.25.766}{\emph{Phys. Rev. D} {\bfseries
  25} (1982) 766}.

\bibitem{Shrock:1982sc}
R.~E. Shrock, \emph{{Electromagnetic Properties and Decays of Dirac and
  Majorana Neutrinos in a General Class of Gauge Theories}},
  \href{https://doi.org/10.1016/0550-3213(82)90273-5}{\emph{Nucl. Phys. B}
  {\bfseries 206} (1982) 359}.

\bibitem{Giunti:2008ve}
C.~Giunti and A.~Studenikin, \emph{{Neutrino electromagnetic properties}},
  \href{https://doi.org/10.1134/S1063778809120126}{\emph{Phys. Atom. Nucl.}
  {\bfseries 72} (2009) 2089}
  [\href{https://arxiv.org/abs/0812.3646}{{\ttfamily 0812.3646}}].

\bibitem{Nieves:1981zt}
J.~F. Nieves, \emph{{Electromagnetic Properties of Majorana Neutrinos}},
  \href{https://doi.org/10.1103/PhysRevD.26.3152}{\emph{Phys. Rev. D}
  {\bfseries 26} (1982) 3152}.

\bibitem{Kayser:1982br}
B.~Kayser, \emph{{Majorana Neutrinos and their Electromagnetic Properties}},
  \href{https://doi.org/10.1103/PhysRevD.26.1662}{\emph{Phys. Rev. D}
  {\bfseries 26} (1982) 1662}.

\bibitem{Aronson:1969ltq}
H.~Aronson, \emph{{Spin-1 electrodynamics with an electric quadrupole moment}},
  \href{https://doi.org/10.1103/PhysRev.186.1434}{\emph{Phys. Rev.} {\bfseries
  186} (1969) 1434}.

\bibitem{Gaemers:1978hg}
K.~J.~F. Gaemers and G.~J. Gounaris, \emph{{Polarization Amplitudes for e+ e-
  ---\ensuremath{>} W+ W- and e+ e- ---\ensuremath{>} Z Z}},
  \href{https://doi.org/10.1007/BF01440226}{\emph{Z. Phys. C} {\bfseries 1}
  (1979) 259}.

\bibitem{Hagiwara:1986vm}
K.~Hagiwara, R.~D. Peccei, D.~Zeppenfeld and K.~Hikasa, \emph{{Probing the Weak
  Boson Sector in e+ e- ---\ensuremath{>} W+ W-}},
  \href{https://doi.org/10.1016/0550-3213(87)90685-7}{\emph{Nucl. Phys. B}
  {\bfseries 282} (1987) 253}.

\bibitem{Pospelov:2000bq}
M.~Pospelov and T.~ter Veldhuis, \emph{{Direct and indirect limits on the
  electromagnetic form-factors of WIMPs}},
  \href{https://doi.org/10.1016/S0370-2693(00)00358-0}{\emph{Phys. Lett. B}
  {\bfseries 480} (2000) 181}
  [\href{https://arxiv.org/abs/hep-ph/0003010}{{\ttfamily hep-ph/0003010}}].

\bibitem{Sigurdson:2004zp}
K.~Sigurdson, M.~Doran, A.~Kurylov, R.~R. Caldwell and M.~Kamionkowski,
  \emph{{Dark-matter electric and magnetic dipole moments}},
  \href{https://doi.org/10.1103/PhysRevD.70.083501}{\emph{Phys. Rev. D}
  {\bfseries 70} (2004) 083501}
  [\href{https://arxiv.org/abs/astro-ph/0406355}{{\ttfamily
  astro-ph/0406355}}].

\bibitem{Masso:2009mu}
E.~Masso, S.~Mohanty and S.~Rao, \emph{{Dipolar Dark Matter}},
  \href{https://doi.org/10.1103/PhysRevD.80.036009}{\emph{Phys. Rev. D}
  {\bfseries 80} (2009) 036009}
  [\href{https://arxiv.org/abs/0906.1979}{{\ttfamily 0906.1979}}].

\bibitem{Chang:2010en}
S.~Chang, N.~Weiner and I.~Yavin, \emph{{Magnetic Inelastic Dark Matter}},
  \href{https://doi.org/10.1103/PhysRevD.82.125011}{\emph{Phys. Rev. D}
  {\bfseries 82} (2010) 125011}
  [\href{https://arxiv.org/abs/1007.4200}{{\ttfamily 1007.4200}}].

\bibitem{Barger:2010gv}
V.~Barger, W.-Y. Keung and D.~Marfatia, \emph{{Electromagnetic properties of
  dark matter: Dipole moments and charge form factor}},
  \href{https://doi.org/10.1016/j.physletb.2010.12.008}{\emph{Phys. Lett. B}
  {\bfseries 696} (2011) 74} [\href{https://arxiv.org/abs/1007.4345}{{\ttfamily
  1007.4345}}].

\bibitem{Banks:2010eh}
T.~Banks, J.-F. Fortin and S.~Thomas, \emph{{Direct Detection of Dark Matter
  Electromagnetic Dipole Moments}},
  \href{https://arxiv.org/abs/1007.5515}{{\ttfamily 1007.5515}}.

\bibitem{Ho:2012bg}
C.~M. Ho and R.~J. Scherrer, \emph{{Anapole Dark Matter}},
  \href{https://doi.org/10.1016/j.physletb.2013.04.039}{\emph{Phys. Lett. B}
  {\bfseries 722} (2013) 341}
  [\href{https://arxiv.org/abs/1211.0503}{{\ttfamily 1211.0503}}].

\bibitem{DelNobile:2012tx}
E.~Del~Nobile, C.~Kouvaris, P.~Panci, F.~Sannino and J.~Virkajarvi,
  \emph{{Light Magnetic Dark Matter in Direct Detection Searches}},
  \href{https://doi.org/10.1088/1475-7516/2012/08/010}{\emph{JCAP} {\bfseries
  08} (2012) 010} [\href{https://arxiv.org/abs/1203.6652}{{\ttfamily
  1203.6652}}].

\bibitem{Kopp:2014tsa}
J.~Kopp, L.~Michaels and J.~Smirnov, \emph{{Loopy Constraints on Leptophilic
  Dark Matter and Internal Bremsstrahlung}},
  \href{https://doi.org/10.1088/1475-7516/2014/04/022}{\emph{JCAP} {\bfseries
  04} (2014) 022} [\href{https://arxiv.org/abs/1401.6457}{{\ttfamily
  1401.6457}}].

\bibitem{Ibarra:2015fqa}
A.~Ibarra and S.~Wild, \emph{{Dirac dark matter with a charged mediator: a
  comprehensive one-loop analysis of the direct detection phenomenology}},
  \href{https://doi.org/10.1088/1475-7516/2015/05/047}{\emph{JCAP} {\bfseries
  05} (2015) 047} [\href{https://arxiv.org/abs/1503.03382}{{\ttfamily
  1503.03382}}].

\bibitem{Hambye:2021xvd}
T.~Hambye and X.-J. Xu, \emph{{Dark matter electromagnetic dipoles: the WIMP
  expectation}},  \href{https://arxiv.org/abs/2106.01403}{{\ttfamily
  2106.01403}}.

\bibitem{Hisano:2020qkq}
J.~Hisano, A.~Ibarra and R.~Nagai, \emph{{Direct detection of vector dark
  matter through electromagnetic multipoles}},
  \href{https://doi.org/10.1088/1475-7516/2020/10/015}{\emph{JCAP} {\bfseries
  10} (2020) 015} [\href{https://arxiv.org/abs/2007.03216}{{\ttfamily
  2007.03216}}].

\bibitem{DelNobile:2014eta}
E.~Del~Nobile, G.~B. Gelmini, P.~Gondolo and J.-H. Huh, \emph{{Direct detection
  of Light Anapole and Magnetic Dipole DM}},
  \href{https://doi.org/10.1088/1475-7516/2014/06/002}{\emph{JCAP} {\bfseries
  06} (2014) 002} [\href{https://arxiv.org/abs/1401.4508}{{\ttfamily
  1401.4508}}].

\bibitem{Garny:2015wea}
M.~Garny, A.~Ibarra and S.~Vogl, \emph{{Signatures of Majorana dark matter with
  t-channel mediators}},
  \href{https://doi.org/10.1142/S0218271815300190}{\emph{Int. J. Mod. Phys. D}
  {\bfseries 24} (2015) 1530019}
  [\href{https://arxiv.org/abs/1503.01500}{{\ttfamily 1503.01500}}].

\bibitem{Cabral-Rosetti:2015cxa}
L.~G. Cabral-Rosetti, M.~Mondrag\'on and E.~Reyes-P\'erez, \emph{{Anapole
  moment of the lightest neutralino in the cMSSM}},
  \href{https://doi.org/10.1016/j.nuclphysb.2016.03.025}{\emph{Nucl. Phys. B}
  {\bfseries 907} (2016) 1} [\href{https://arxiv.org/abs/1504.01213}{{\ttfamily
  1504.01213}}].

\bibitem{Abe:2018emu}
T.~Abe, M.~Fujiwara and J.~Hisano, \emph{{Loop corrections to dark matter
  direct detection in a pseudoscalar mediator dark matter model}},
  \href{https://doi.org/10.1007/JHEP02(2019)028}{\emph{JHEP} {\bfseries 02}
  (2019) 028} [\href{https://arxiv.org/abs/1810.01039}{{\ttfamily
  1810.01039}}].

\bibitem{Sandick:2016zut}
P.~Sandick, K.~Sinha and F.~Teng, \emph{{Simplified Dark Matter Models with
  Charged Mediators: Prospects for Direct Detection}},
  \href{https://doi.org/10.1007/JHEP10(2016)018}{\emph{JHEP} {\bfseries 10}
  (2016) 018} [\href{https://arxiv.org/abs/1608.00642}{{\ttfamily
  1608.00642}}].

\bibitem{Kang:2018oej}
S.~Kang, S.~Scopel, G.~Tomar, J.-H. Yoon and P.~Gondolo, \emph{{Anapole Dark
  Matter after DAMA/LIBRA-phase2}},
  \href{https://doi.org/10.1088/1475-7516/2018/11/040}{\emph{JCAP} {\bfseries
  11} (2018) 040} [\href{https://arxiv.org/abs/1808.04112}{{\ttfamily
  1808.04112}}].

\bibitem{Baker:2018uox}
M.~J. Baker and A.~Thamm, \emph{{Leptonic WIMP Coannihilation and the Current
  Dark Matter Search Strategy}},
  \href{https://doi.org/10.1007/JHEP10(2018)187}{\emph{JHEP} {\bfseries 10}
  (2018) 187} [\href{https://arxiv.org/abs/1806.07896}{{\ttfamily
  1806.07896}}].

\bibitem{Arina:2020mxo}
C.~Arina, A.~Cheek, K.~Mimasu and L.~Pagani, \emph{{Light and Darkness:
  consistently coupling dark matter to photons via effective operators}},
  \href{https://doi.org/10.1140/epjc/s10052-021-09010-1}{\emph{Eur. Phys. J. C}
  {\bfseries 81} (2021) 223}
  [\href{https://arxiv.org/abs/2005.12789}{{\ttfamily 2005.12789}}].

\bibitem{Kuo:2021mtp}
J.-L. Kuo, M.~Pospelov and J.~Pradler, \emph{{Terrestrial probes of
  electromagnetically interacting dark radiation}},
  \href{https://doi.org/10.1103/PhysRevD.103.115030}{\emph{Phys. Rev. D}
  {\bfseries 103} (2021) 115030}
  [\href{https://arxiv.org/abs/2102.08409}{{\ttfamily 2102.08409}}].

\bibitem{Bardeen:1972vi}
W.~A. Bardeen, R.~Gastmans and B.~E. Lautrup, \emph{{Static quantities in
  Weinberg's model of weak and electromagnetic interactions}},
  \href{https://doi.org/10.1016/0550-3213(72)90218-0}{\emph{Nucl. Phys. B}
  {\bfseries 46} (1972) 319}.

\bibitem{Abers:1973qs}
E.~S. Abers and B.~W. Lee, \emph{{Gauge Theories}},
  \href{https://doi.org/10.1016/0370-1573(73)90027-6}{\emph{Phys. Rept.}
  {\bfseries 9} (1973) 1}.

\bibitem{Fujikawa:1972fe}
K.~Fujikawa, B.~W. Lee and A.~I. Sanda, \emph{Generalized renormalizable gauge
  formulation of spontaneously broken gauge theories},
  \href{https://doi.org/10.1103/PhysRevD.6.2923}{\emph{Phys. Rev. D} {\bfseries
  6} (1972) 2923}.

\bibitem{Lee:1973fw}
S.~Y. Lee, \emph{{Higher-order corrections to leptonic processes and the
  renormalization of weinberg's theory of weak interactions in the unitary
  gauge}}, \href{https://doi.org/10.1103/PhysRevD.6.1701}{\emph{Phys. Rev. D}
  {\bfseries 6} (1972) 1701}.

\bibitem{PhysRevD.43.2956}
M.~J. Musolf and B.~R. Holstein, \emph{Observability of the anapole moment and
  neutrino charge radius},
  \href{https://doi.org/10.1103/PhysRevD.43.2956}{\emph{Phys. Rev. D}
  {\bfseries 43} (1991) 2956}.

\bibitem{Papavassiliou:1993qe}
J.~Papavassiliou and C.~Parrinello, \emph{{Gauge invariant top quark
  form-factors from e+ e- experiments}},
  \href{https://doi.org/10.1103/PhysRevD.50.3059}{\emph{Phys. Rev. D}
  {\bfseries 50} (1994) 3059}
  [\href{https://arxiv.org/abs/hep-ph/9311284}{{\ttfamily hep-ph/9311284}}].

\bibitem{CabralRosetti:2002tf}
L.~G. Cabral-Rosetti, G.~Lopez~Castro and J.~Pestieau, \emph{{One loop
  electroweak corrections to the muon anomalous magnetic moment using the pinch
  technique}},  \href{https://arxiv.org/abs/hep-ph/0211437}{{\ttfamily
  hep-ph/0211437}}.

\bibitem{Bernabeu:2007rr}
J.~Bernabeu, G.~A. Gonzalez-Sprinberg, J.~Papavassiliou and J.~Vidal,
  \emph{{Tau anomalous magnetic moment form-factor at super B/flavor
  factories}},
  \href{https://doi.org/10.1016/j.nuclphysb.2007.09.001}{\emph{Nucl. Phys. B}
  {\bfseries 790} (2008) 160}
  [\href{https://arxiv.org/abs/0707.2496}{{\ttfamily 0707.2496}}].

\bibitem{Cornwall:1989gv}
J.~M. Cornwall and J.~Papavassiliou, \emph{{Gauge Invariant Three Gluon Vertex
  in QCD}}, \href{https://doi.org/10.1103/PhysRevD.40.3474}{\emph{Phys. Rev. D}
  {\bfseries 40} (1989) 3474}.

\bibitem{Papavassiliou:1989zd}
J.~Papavassiliou, \emph{{Gauge Invariant Proper Selfenergies and Vertices in
  Gauge Theories with Broken Symmetry}},
  \href{https://doi.org/10.1103/PhysRevD.41.3179}{\emph{Phys. Rev. D}
  {\bfseries 41} (1990) 3179}.

\bibitem{Bernabeu:2000hf}
J.~Bernabeu, L.~G. Cabral-Rosetti, J.~Papavassiliou and J.~Vidal, \emph{{On the
  charge radius of the neutrino}},
  \href{https://doi.org/10.1103/PhysRevD.62.113012}{\emph{Phys. Rev. D}
  {\bfseries 62} (2000) 113012}
  [\href{https://arxiv.org/abs/hep-ph/0008114}{{\ttfamily hep-ph/0008114}}].

\bibitem{PhysRevD.61.013001}
A.~Rosado, \emph{Physical electroweak anapole moment for the neutrino},
  \href{https://doi.org/10.1103/PhysRevD.61.013001}{\emph{Phys. Rev. D}
  {\bfseries 61} (1999) 013001}.

\bibitem{Bernabeu:2002pd}
J.~Bernabeu, J.~Papavassiliou and J.~Vidal, \emph{{The Neutrino charge radius
  is a physical observable}},
  \href{https://doi.org/10.1016/j.nuclphysb.2003.12.025}{\emph{Nucl. Phys. B}
  {\bfseries 680} (2004) 450}
  [\href{https://arxiv.org/abs/hep-ph/0210055}{{\ttfamily hep-ph/0210055}}].

\bibitem{Binosi:2009qm}
D.~Binosi and J.~Papavassiliou, \emph{{Pinch Technique: Theory and
  Applications}},
  \href{https://doi.org/10.1016/j.physrep.2009.05.001}{\emph{Phys. Rept.}
  {\bfseries 479} (2009) 1} [\href{https://arxiv.org/abs/0909.2536}{{\ttfamily
  0909.2536}}].

\bibitem{Denner:1994nn}
A.~Denner, G.~Weiglein and S.~Dittmaier, \emph{{Gauge invariance of Green
  functions: Background field method versus pinch technique}},
  \href{https://doi.org/10.1016/0370-2693(94)90162-7}{\emph{Phys. Lett. B}
  {\bfseries 333} (1994) 420}
  [\href{https://arxiv.org/abs/hep-ph/9406204}{{\ttfamily hep-ph/9406204}}].

\bibitem{Hashimoto:1994ct}
S.~Hashimoto, J.~Kodaira, Y.~Yasui and K.~Sasaki, \emph{{The Background field
  method: Alternative way of deriving the pinch technique's results}},
  \href{https://doi.org/10.1103/PhysRevD.50.7066}{\emph{Phys. Rev. D}
  {\bfseries 50} (1994) 7066}
  [\href{https://arxiv.org/abs/hep-ph/9406271}{{\ttfamily hep-ph/9406271}}].

\bibitem{Papavassiliou:1994yi}
J.~Papavassiliou, \emph{{On the connection between the pinch technique and the
  background field method}},
  \href{https://doi.org/10.1103/PhysRevD.51.856}{\emph{Phys. Rev. D} {\bfseries
  51} (1995) 856} [\href{https://arxiv.org/abs/hep-ph/9410385}{{\ttfamily
  hep-ph/9410385}}].

\bibitem{Denner:1994xt}
A.~Denner, G.~Weiglein and S.~Dittmaier, \emph{{Application of the background
  field method to the electroweak standard model}},
  \href{https://doi.org/10.1016/0550-3213(95)00037-S}{\emph{Nucl. Phys. B}
  {\bfseries 440} (1995) 95}
  [\href{https://arxiv.org/abs/hep-ph/9410338}{{\ttfamily hep-ph/9410338}}].

\bibitem{Denner:1992vzaFeynmanRulesMajorana1}
A.~Denner, H.~Eck, O.~Hahn and J.~Kublbeck, \emph{{Feynman rules for fermion
  number violating interactions}},
  \href{https://doi.org/10.1016/0550-3213(92)90169-C}{\emph{Nucl. Phys. B}
  {\bfseries 387} (1992) 467}.

\bibitem{Denner:1992meFeynmanRulesMajorana2}
A.~Denner, H.~Eck, O.~Hahn and J.~Kublbeck, \emph{{Compact Feynman rules for
  Majorana fermions}},
  \href{https://doi.org/10.1016/0370-2693(92)91045-B}{\emph{Phys. Lett. B}
  {\bfseries 291} (1992) 278}.

\bibitem{MERTIG1991345}
R.~Mertig, M.~Böhm and A.~Denner, \emph{Feyn calc - computer-algebraic
  calculation of feynman amplitudes},
  \href{https://doi.org/https://doi.org/10.1016/0010-4655(91)90130-D}{\emph{Computer
  Physics Communications} {\bfseries 64} (1991) 345}.

\bibitem{Shtabovenko:2016sxi}
V.~Shtabovenko, R.~Mertig and F.~Orellana, \emph{{New Developments in FeynCalc
  9.0}}, \href{https://doi.org/10.1016/j.cpc.2016.06.008}{\emph{Comput. Phys.
  Commun.} {\bfseries 207} (2016) 432}
  [\href{https://arxiv.org/abs/1601.01167}{{\ttfamily 1601.01167}}].

\bibitem{Shtabovenko:2020gxv}
V.~Shtabovenko, R.~Mertig and F.~Orellana, \emph{{FeynCalc 9.3: New features
  and improvements}},
  \href{https://doi.org/10.1016/j.cpc.2020.107478}{\emph{Comput. Phys. Commun.}
  {\bfseries 256} (2020) 107478}
  [\href{https://arxiv.org/abs/2001.04407}{{\ttfamily 2001.04407}}].

\bibitem{Shtabovenko:2016whf}
V.~Shtabovenko, \emph{{FeynHelpers: Connecting FeynCalc to FIRE and
  Package-X}}, \href{https://doi.org/10.1016/j.cpc.2017.04.014}{\emph{Comput.
  Phys. Commun.} {\bfseries 218} (2017) 48}
  [\href{https://arxiv.org/abs/1611.06793}{{\ttfamily 1611.06793}}].

\bibitem{Patel:2015tea}
H.~H. Patel, \emph{{Package-X: A Mathematica package for the analytic
  calculation of one-loop integrals}},
  \href{https://doi.org/10.1016/j.cpc.2015.08.017}{\emph{Comput. Phys. Commun.}
  {\bfseries 197} (2015) 276}
  [\href{https://arxiv.org/abs/1503.01469}{{\ttfamily 1503.01469}}].

\bibitem{Griest:1990kh}
K.~Griest and D.~Seckel, \emph{{Three exceptions in the calculation of relic
  abundances}}, \href{https://doi.org/10.1103/PhysRevD.43.3191}{\emph{Phys.
  Rev. D} {\bfseries 43} (1991) 3191}.

\bibitem{Baker:2015qna}
M.~J. Baker et~al., \emph{{The Coannihilation Codex}},
  \href{https://doi.org/10.1007/JHEP12(2015)120}{\emph{JHEP} {\bfseries 12}
  (2015) 120} [\href{https://arxiv.org/abs/1510.03434}{{\ttfamily
  1510.03434}}].

\bibitem{Flores:1989ru}
R.~Flores, K.~A. Olive and S.~Rudaz, \emph{{Radiative Processes in Lsp
  Annihilation}},
  \href{https://doi.org/10.1016/0370-2693(89)90760-0}{\emph{Phys. Lett. B}
  {\bfseries 232} (1989) 377}.

\bibitem{Garny:2011ii}
M.~Garny, A.~Ibarra and S.~Vogl, \emph{{Dark matter annihilations into two
  light fermions and one gauge boson: General analysis and antiproton
  constraints}},
  \href{https://doi.org/10.1088/1475-7516/2012/04/033}{\emph{JCAP} {\bfseries
  04} (2012) 033} [\href{https://arxiv.org/abs/1112.5155}{{\ttfamily
  1112.5155}}].

\bibitem{Martin:2007gf}
S.~P. Martin, \emph{{Compressed supersymmetry and natural neutralino dark
  matter from top squark-mediated annihilation to top quarks}},
  \href{https://doi.org/10.1103/PhysRevD.75.115005}{\emph{Phys. Rev. D}
  {\bfseries 75} (2007) 115005}
  [\href{https://arxiv.org/abs/hep-ph/0703097}{{\ttfamily hep-ph/0703097}}].

\bibitem{Dreiner:2012gx}
H.~K. Dreiner, M.~Kramer and J.~Tattersall, \emph{{How low can SUSY go?
  Matching, monojets and compressed spectra}},
  \href{https://doi.org/10.1209/0295-5075/99/61001}{\emph{EPL} {\bfseries 99}
  (2012) 61001} [\href{https://arxiv.org/abs/1207.1613}{{\ttfamily
  1207.1613}}].

\bibitem{ATLAS:2019gti}
{\scshape ATLAS} collaboration, \emph{{Search for direct stau production in
  events with two hadronic $\tau$-leptons in $\sqrt{s} = 13$ TeV $pp$
  collisions with the ATLAS detector}},
  \href{https://doi.org/10.1103/PhysRevD.101.032009}{\emph{Phys. Rev. D}
  {\bfseries 101} (2020) 032009}
  [\href{https://arxiv.org/abs/1911.06660}{{\ttfamily 1911.06660}}].

\bibitem{Berggren:2001kb}
M.~Berggren, \emph{{Stau searches at LEP}},
  \href{https://doi.org/10.1016/S0920-5632(01)01247-6}{\emph{Nucl. Phys. B
  Proc. Suppl.} {\bfseries 98} (2001) 342}.

\bibitem{Allanach:2017hcf}
B.~C. Allanach and T.~Cridge, \emph{{The Calculation of Sparticle and Higgs
  Decays in the Minimal and Next-to-Minimal Supersymmetric Standard Models:
  SOFTSUSY4.0}}, \href{https://doi.org/10.1016/j.cpc.2017.07.021}{\emph{Comput.
  Phys. Commun.} {\bfseries 220} (2017) 417}
  [\href{https://arxiv.org/abs/1703.09717}{{\ttfamily 1703.09717}}].

\bibitem{Belanger:2013oya}
G.~Belanger, F.~Boudjema, A.~Pukhov and A.~Semenov, \emph{{micrOMEGAs$\_$3: A
  program for calculating dark matter observables}},
  \href{https://doi.org/10.1016/j.cpc.2013.10.016}{\emph{Comput. Phys. Commun.}
  {\bfseries 185} (2014) 960}
  [\href{https://arxiv.org/abs/1305.0237}{{\ttfamily 1305.0237}}].

\bibitem{Alguero:2021dig}
G.~Alguero, J.~Heisig, C.~Khosa, S.~Kraml, S.~Kulkarni, A.~Lessa et~al.,
  \emph{{Constraining new physics with SModelS version 2}},
  \href{https://arxiv.org/abs/2112.00769}{{\ttfamily 2112.00769}}.

\bibitem{Bechtle:2013wla}
P.~Bechtle, O.~Brein, S.~Heinemeyer, O.~Stal, T.~Stefaniak, G.~Weiglein et~al.,
  \emph{{HiggsBounds-4: Improved Tests of Extended Higgs Sectors against
  Exclusion Bounds from LEP, the Tevatron and the LHC}},
  \href{https://doi.org/10.1140/epjc/s10052-013-2693-2}{\emph{Eur. Phys. J. C}
  {\bfseries 74} (2014) 2693}
  [\href{https://arxiv.org/abs/1311.0055}{{\ttfamily 1311.0055}}].

\bibitem{Stal:2013hwa}
O.~Stal and T.~Stefaniak, \emph{{Constraining extended Higgs sectors with
  HiggsSignals}}, \href{https://doi.org/10.22323/1.180.0314}{\emph{PoS}
  {\bfseries EPS-HEP2013} (2013) 314}
  [\href{https://arxiv.org/abs/1310.4039}{{\ttfamily 1310.4039}}].

\bibitem{Mahmoudi:2009zz}
F.~Mahmoudi, \emph{{SuperIso v3.0, flavor physics observables calculations:
  Extension to NMSSM}},
  \href{https://doi.org/10.1016/j.cpc.2009.05.001}{\emph{Comput. Phys. Commun.}
  {\bfseries 180} (2009) 1718}.

\bibitem{Athron:2015rva}
P.~Athron, M.~Bach, H.~G. Fargnoli, C.~Gnendiger, R.~Greifenhagen, J.-h. Park
  et~al., \emph{{GM2Calc: Precise MSSM prediction for $(g - 2)$ of the muon}},
  \href{https://doi.org/10.1140/epjc/s10052-015-3870-2}{\emph{Eur. Phys. J. C}
  {\bfseries 76} (2016) 62} [\href{https://arxiv.org/abs/1510.08071}{{\ttfamily
  1510.08071}}].

\bibitem{Buckley:2013jua}
A.~Buckley, \emph{{PySLHA: a Pythonic interface to SUSY Les Houches Accord
  data}}, \href{https://doi.org/10.1140/epjc/s10052-015-3638-8}{\emph{Eur.
  Phys. J. C} {\bfseries 75} (2015) 467}
  [\href{https://arxiv.org/abs/1305.4194}{{\ttfamily 1305.4194}}].

\bibitem{Allanach:2008qq}
B.~C. Allanach et~al., \emph{{SUSY Les Houches Accord 2}},
  \href{https://doi.org/10.1016/j.cpc.2008.08.004}{\emph{Comput. Phys. Commun.}
  {\bfseries 180} (2009) 8} [\href{https://arxiv.org/abs/0801.0045}{{\ttfamily
  0801.0045}}].

\bibitem{Helm:1956zz}
R.~H. Helm, \emph{{Inelastic and Elastic Scattering of 187-Mev Electrons from
  Selected Even-Even Nuclei}},
  \href{https://doi.org/10.1103/PhysRev.104.1466}{\emph{Phys. Rev.} {\bfseries
  104} (1956) 1466}.

\bibitem{Lewin:1995rx}
J.~Lewin and P.~Smith, \emph{{Review of mathematics, numerical factors, and
  corrections for dark matter experiments based on elastic nuclear recoil}},
  \href{https://doi.org/10.1016/S0927-6505(96)00047-3}{\emph{Astropart. Phys.}
  {\bfseries 6} (1996) 87}.

\bibitem{Aprile:2018dbl}
{\scshape XENON} collaboration, \emph{{Dark Matter Search Results from a One
  Ton-Year Exposure of XENON1T}},
  \href{https://doi.org/10.1103/PhysRevLett.121.111302}{\emph{Phys. Rev. Lett.}
  {\bfseries 121} (2018) 111302}
  [\href{https://arxiv.org/abs/1805.12562}{{\ttfamily 1805.12562}}].

\bibitem{Agnese:2014aze}
{\scshape SuperCDMS} collaboration, \emph{{Search for Low-Mass Weakly
  Interacting Massive Particles with SuperCDMS}},
  \href{https://doi.org/10.1103/PhysRevLett.112.241302}{\emph{Phys. Rev. Lett.}
  {\bfseries 112} (2014) 241302}
  [\href{https://arxiv.org/abs/1402.7137}{{\ttfamily 1402.7137}}].

\bibitem{Amole:2019fdf}
{\scshape PICO} collaboration, \emph{{Dark Matter Search Results from the
  Complete Exposure of the PICO-60 C$_3$F$_8$ Bubble Chamber}},
  \href{https://doi.org/10.1103/PhysRevD.100.022001}{\emph{Phys. Rev. D}
  {\bfseries 100} (2019) 022001}
  [\href{https://arxiv.org/abs/1902.04031}{{\ttfamily 1902.04031}}].

\bibitem{Aprile:2015uzo}
{\scshape XENON} collaboration, \emph{{Physics reach of the XENON1T dark matter
  experiment}},
  \href{https://doi.org/10.1088/1475-7516/2016/04/027}{\emph{JCAP} {\bfseries
  04} (2016) 027} [\href{https://arxiv.org/abs/1512.07501}{{\ttfamily
  1512.07501}}].

\bibitem{ELLIS2017103}
J.~P. Ellis, \emph{Tikz-feynman: Feynman diagrams with tikz},
  \href{https://doi.org/https://doi.org/10.1016/j.cpc.2016.08.019}{\emph{Computer
  Physics Communications} {\bfseries 210} (2017) 103}.

\bibitem{Haber:1984rc}
H.~E. Haber and G.~L. Kane, \emph{{The Search for Supersymmetry: Probing
  Physics Beyond the Standard Model}},
  \href{https://doi.org/10.1016/0370-1573(85)90051-1}{\emph{Phys. Rept.}
  {\bfseries 117} (1985) 75}.

\bibitem{Drees:2004SUSY}
M.~Drees, R.~Godbole and P.~Roy, \emph{Theory and Phenomenology of Sparticles:
  An Account of Four-Dimensional N =1 Supersymmetry in High Energy Physics}.
  World Scientific, 01, 2005,
  \href{https://doi.org/10.1142/4001}{10.1142/4001}.

\end{thebibliography}\endgroup
\end{document}